\documentclass{aa}

\usepackage[T1]{fontenc}
\usepackage{graphicx}
\usepackage{amsmath}
\usepackage{multicol,multirow}
\usepackage{tabularx}
\usepackage{booktabs} 
\usepackage{color,soul}
\usepackage{xspace} 
\usepackage{xcolor}
\usepackage{subcaption}
\usepackage[flushleft]{threeparttable}

\usepackage{txfonts}

\usepackage{natbib}
\usepackage[colorlinks=true,linkcolor=blue,citecolor=blue]{hyperref}
\bibpunct{(}{)}{;}{a}{}{,}

\newcommand{\chandra}{\textit{Chandra}\xspace}
\newcommand{\rxte}{\textit{RXTE}\xspace}

\newcommand{\ixpe}{{IXPE}\xspace}
\newcommand{\integral}{\textit{INTEGRAL}\xspace}

\newcommand{\nicer}{{NICER}\xspace }
\newcommand{\nustar}{\textit{NuSTAR}\xspace}

\newcommand{\lumcgs}{erg~s$^{-1}$\xspace}

\xspaceaddexceptions{+ - *}

\graphicspath{{./}{Figures/}}

\begin{document}

\title{X-ray polarization of Z-type neutron star low-mass X-ray binaries}
\subtitle{I. Model-independent, time-resolved X-ray polarimetry}
\titlerunning{Polarization of Z-Sources I.}
\authorrunning{A. Gnarini et al.}

\author{Andrea Gnarini \inst{\ref{in:UniRoma3}}\fnmsep\thanks{E-mail: \href{mailto:andrea.gnarini@uniroma3.it}{{\tt andrea.gnarini@uniroma3.it}}}
\and Francesco Ursini \inst{\ref{in:UniRoma3}}
\and Giorgio Matt \inst{\ref{in:UniRoma3}}
\and Stefano Bianchi \inst{\ref{in:UniRoma3}}
\and Fiamma Capitanio \inst{\ref{in:INAF-IAPS}}
\and Massimo Cocchi \inst{\ref{in:INAF-OAC}}
\and Sergio Fabiani \inst{\ref{in:INAF-IAPS}}
\and Ruben Farinelli \inst{\ref{in:INAF-OAS}}
\and Antonella Tarana \inst{\ref{in:INAF-IAPS}}
}

\institute{Dipartimento di Matematica e Fisica, Università degli Studi Roma Tre, Via della Vasca Navale 84, I-00146 Roma, Italy \label{in:UniRoma3}
\and
INAF -- Istituto di Astrofisica e Planetologia Spaziali, Via del Fosso del Cavaliere 100, I-00133 Roma, Italy \label{in:INAF-IAPS}
\and
INAF -- Osservatorio Astronomico di Cagliari, via della Scienza 5, I-09047 Selargius (CA), Italy \label{in:INAF-OAC}
\and
INAF -- Osservatorio di Astrofisica e Scienza dello Spazio, Via P. Gobetti 101, I-40129 Bologna, Italy \label{in:INAF-OAS}
}

\date{Received XXX; accepted YYY}


\abstract{ 
    Z-sources are a particular class of neutron star low-mass X-ray binaries characterized by a wide Z-like track in their hard color-soft color (or hardness-intensity) diagrams, with three branches: the horizontal (HB), the normal (NB), and the flaring branch (FB). Spectropolarimetric observations with the Imaging X-ray Polarimetry Explorer (\ixpe) show that the polarization in these sources varies along the Z-track, reaching unexpectedly high values in the HB. In this work, we collected all the polarimetric results obtained so far from observations of Z-sources with \ixpe, using a model-independent analysis with \textsc{ixpeobssim}. We first performed a detailed characterization of the spectral state of each source along the Z-track using \ixpe, along with the Nuclear Spectroscopic Telescope Array (\nustar) and the Neutron Star Interior Composition Explorer (\nicer) data and then estimated the polarization for each branch. Although we confirm that the average polarization in the 2--8 keV band decreases moving from the HB to the NB for all three Z-sources observed in these branches, we also observe a qualitatively increasing trend from the NB to the FB. Whereas this increase is clearly significant for Cyg~X-2 and Sco~X-1, the polarization remains consistent at the 90\% confidence level for GX~5--1 and GX~349+2, while for XTE~J1701--462 and GX~340+0 only upper limits are found in the FB. For most sources, the average polarization angle in the 2--8 keV range remains consistent along the CCD; however, we observe a significant rotation for both Sco~X-1 and GX~349+2 (at the 90\% confidence level) as they move from the NB to the FB. In addition, we observe a significant increase in the polarization degree with energy in most of the observed Z-sources, with some also exhibiting a rotation of the polarization angle with energy (approximately by $20\degr-30\degr$).
}

\keywords{accretion, accretion disks -- stars: neutron -- X-rays: binaries -- polarization}

\maketitle


\section{Introduction}

The launch of the NASA and Italian Space Agency Imaging X-ray Polarimetry Explorer (\ixpe; \citealt{Weisskopf.2022}) on 2021 December 9 opened a new window in X-ray astronomy, providing for the first time space-, time-, and energy-resolved polarimetric observations of different classes of X-ray sources. Among the brightest X-ray sources, weakly magnetized neutron stars in low-mass X-ray binaries (NS-LMXBs) represent ideal targets for studying radiation processes in the strong-gravity regime. NS-LMXBs accrete matter via Roche-lobe overflow, typically from a late main-sequence star. Following their tracks in hard-color/soft-color diagrams (CCDs) or in hardness-intensity diagrams (HIDs; \citealt{Hasinger.VanDerKlis.1989,VanDerKlis.1989}), NS-LMXBs are traditionally divided into: Atoll sources ($L_{\rm X} \sim 10^{36}$ \lumcgs), which exhibit compact tracks in a rounded single spot in the hard region of the CCD, termed ``island'' state or in a ``banana'' shape for bright Atolls ($L_{\rm X} \sim 10^{37}-10^{38}$ \lumcgs; \citealt{VanDerKlis.1995}); Z-sources ($L_{\rm X} > 10^{38}$ \lumcgs), characterized by a wide Z-like three-branch pattern in the CCD, composed of the horizontal branch (HB), the normal branch (NB) and flaring branch (FB). The luminosity and accretion rate are expected to increase as the sources move from the HB to the FB \citep{Migliari.Fender.2006}. The Z-sources can be further divided into two subgroups: Cyg~X-2-like sources (e.g., Cyg~X-2, GX~5--1, and GX~340+0) display the full Z-track with full HB and NB, but with weak FB; in contrast, Sco~X-1-like sources (e.g., Sco~X-1, GX~17+2, and GX~349+2) show little or no HB, but the FB is stronger and more frequent, with large increases in X-ray intensity \citep{Kuulkers.etAl.1994,Church.etAl.2012}. 

Based on \ixpe observations to date, the Z-sources appear to be the most polarized NS-LMXBs in the 2--8 keV energy band \citep{Farinelli.etAl.2023,Cocchi.etAl.2023,Fabiani.etAl.2024,LaMonaca.etAl.2024,Bhargava.etAl.2024a,Bhargava.etAl.2024b}, reaching unexpectedly high values along the HB ($\approx 4\%$). The observed polarization strongly depends on the position along the CCDs or HIDs: it decreases (up to $\approx$ 1\%) with no significant rotation of the polarization angle \citep{Cocchi.etAl.2023,Fabiani.etAl.2024} while moving from the HB to the NB or FB. However, the polarization along the NB and FB separately has not been resolved, with the exception of Sco~X-1 \citep{LaMonaca.etAl.2024}. In most sources, the observed polarization seems to be related to the hard component: the typical X-ray emission of NS-LMXBs is generally described by a soft thermal component related to the accretion disk \citep{Mitsuda.etAl.1984,Mitsuda.etAl.1989} or the NS surface emission \citep{White.etAl.1988}, plus a harder component produced by inverse Compton scattering of photons in the hot electron corona,. However, the shape and nature of this region are unclear. X-ray polarization strongly depends on the geometry of the Comptonizing region; therefore, by studying the polarization properties of Z-sources along the CCDs or HIDs, we can obtain insights into the nature and evolution of the hot electron plasma corona in these systems.

In this work, we collect the polarimetric data of each Z-source observed by \ixpe to date and perform a new analysis: in this first paper, we present the new model-independent, time-resolved polarimetric analysis using the \texttt{PCUBE} algorithm of \textsc{ixpeobssim} \citep{Baldini.etAl.2022}. We follow the same data reduction procedure and methodological approach for all Z-sources considered, to enable a direct comparison between the different sources and avoid unnecessary biases. The paper is structured as follows. In the next section, we introduce each Z-source observed with \ixpe. In Sect. \ref{sec:Observations}, we report the observations and data reduction used in this work. In Sect. \ref{sec:Timing}, we describe the timing properties of each source to characterize their spectral state. In Sect. \ref{sec:Polarization} and \ref{sec:Results}, we discuss the polarimetric results obtained with \ixpe using a model-independent analysis, followed by the conclusions in Sect. \ref{sec:Conclusions}.

\section{Z-Sources}\label{sec:Sources}

\subsection{Cyg~X-2}

Cyg~X-2 was first identified in X-rays in the 1960s \citep{Byram.etAl.1966}. When the Rossi X-ray Timing Explorer (\rxte) detected a Type-I X-ray burst with the source in a high-intensity state, it was verified to be an NS-LMXB \citep{Smale.1998}.  The source has been observed several times in the optical \citep{vanParadijs.etAl.1990,Casares.etAl.1998,Orosz.Kuulkers.1999}. From the modeling of ellipsoidal light curves, the NS mass is estimated to be 1.71 $\pm$ 0.21 $M_\odot$ for an inclination of 62.5\degr\ $\pm$ 4\degr\ \citep{Orosz.Kuulkers.1999}. The estimated distance of Cyg~X-2 is between 8 and 11 kpc \citep{Cowley.etAl.1979,Smale.1998}, while optical observations tend to locate the source at 7.2 $\pm$ 1.1 kpc \citep{Orosz.Kuulkers.1999}. 

Cyg~X-2 was subsequently classified as a Z-source by \cite{Hasinger.VanDerKlis.1989}, covering the entire Z-track generally within a few days, with irregular dips during flaring activity. Due to its high X-ray flux and persistent nature, the spectra of Cyg~X-2 have been extensively studied. Broadband spectra are typically well described by a combination of a soft multicolor disk emission and a hard Comptonization component in a region with low temperature ($kT_{\rm e} \sim 3$ keV) and relatively high optical depth ($\tau \sim 5-10$, depending on the geometry). In particular, as Cyg~X-2 moves from the HB to the NB, the inner edge of the accretion disk appears to be closer to the NS surface as the accretion rate increases \citep{DiSalvo.etAl.2002}. In addition, features due to reflection off the accretion disk have also been detected \citep{Smale.etAl.1993,Cackett.etAl.2010,Mondal.etAl.2018,Ludlam.2024}. Considering a full reflection model with recent observations with \nicer and \nustar, \cite{Ludlam.etAl.2022} estimate the inclination of the system to be $i \sim 60-70$\degr, consistent with the optical results \citep{Orosz.Kuulkers.1999}, while the inner disk appears to remain close to the innermost stable circular orbit as the source moves along the CCD. Moreover, an emission line at $\sim 1$ keV has also been observed with different missions \citep{Vrtilek.etAl.1986,Kuulkers.etAl.1997,DiSalvo.etAl.2002,Farinelli.etAl.2009,Cackett.etAl.2010}, which is probably due to photoionized material farther out in the disk \citep{Vrtilek.etAl.1986}. Along the HB, a transient hard tail was detected above 30 keV, which appears to be correlated with the radio jet emission \citep{DAmico.etAl.2001,Paizis.etAl.2006}. The presence of a subrelativistic radio jet at $141\degr$ east of north was found with Very-long-baseline interferometry (VLBI) at 5 GHz when the source was in the HB \citep{Spencer.etAl.2013}. 

\subsection{XTE~J1701--462}

After more than 15 years of quiescence since its first outburst \citep{Lin.etAl.2.2009,Lin.etAl.2009}, XTE~J1701--462 became bright again on 2022 September 6 \citep{MAXI.2022,Swift.2022}. Discovered at the beginning of its 2006 outburst, the source was soon identified as the first transient Z-source \citep{Remillard.etAl.2006,Homan.etAl.2006,Homan.etAl.2007a,Homan.etAl.2007b}, also showing kHz quasi-periodic oscillations (QPOs) in the \rxte time domain during its soft state \citep{Sanna.etAl.2010}. Among NS-LMXBs, XTE~J1701--462 is unique, as it is the only known source to exhibit all spectral substates during its outburst evolution, from Cyg~X-2-like to Sco~X-1-like Z-source, down to the bright Atoll, and finally to the island Atoll state \citep{Lin.etAl.2009,Homan.etAl.2010}. Since there are no absorption dips or eclipses in its light curves, the inclination of the system cannot be high ($\lesssim$ 70\degr--75\degr; \citealt{Lin.etAl.2009}).

During its Z-source state, XTE~J1701--462 spectra can be well described by a combination of soft thermal disk emission and a hard Comptonization component \citep{Lin.etAl.2009,Wang.etAl.2014,Cocchi.etAl.2023,Thomas.etAl.2024}. Reflection features have been observed with \nicer and \nustar. Using full reflection models, the inclination of the system appears to be very low ($\approx$ 20\degr--30\degr; \citealt{Thomas.etAl.2024}), with an inner disk radius close to the innermost stable circular orbit ($\approx 1.5-2$ $R_\text{ISCO}$; \citealt{Thomas.etAl.2024}). The typical transient hard tail was also observed during the HB of its first outburst \citep{Paizis.etAl.2006}.

\subsection{GX~5--1}

GX~5--1 is a Z-source discovered in 1968 with an Aerobee rocket \citep{Bradt.etAl.1968}. It is located near the Galactic center at a distance of about 7.5 kpc \citep{Fabiani.etAl.2024}. This estimate was derived from the equivalent hydrogen column density and a 3D extinction map of the $K_s$ radiation in the Milky Way \citep{Gambino.etAl.2016}. This result is consistent with values obtained previously (e.g., \citealt{Penninx.1989,Smith.etAl.2006}). Early GX~5--1 observations were likely contaminated by the nearby black hole LMXB GRS~1758--258, located only 40' away. However, GX~5--1 is much brighter below 20 keV once the two sources were resolved \citep{Sunyaev.etAl.1991,Gilfanov.etAl.1993}. Radio emission from GX~5--1 due to a compact jet was also detected \citep{Fender.Hendry.2000} and a probable infrared companion candidate has been identified thanks to the precise localization of the radio counterpart \citep{Jonker.etAl.2000}. Moreover,  \chandra observations identified an X-ray halo related to scattering of X-ray photons, arising from the presence of multiple clouds along the line of sight \citep{Smith.etAl.2006,Clark.2018}. 

Unlike most NS-LMXBs, no reflection features have been observed in any of GX~5--1 spectra \citep{Homan.etAl.2018,Fabiani.etAl.2024}. The X-ray continuum can be well described by a combination of multicolor disk blackbody emission and Comptonization of blackbody photons emitted from the NS surface in the boundary or spreading layer. The lack of reflection may be due to a highly ionized accretion disk, for which the Fe K$\alpha$ line becomes broader and weaker as a result of Compton scattering \citep{Ross.etAl.1999,Homan.etAl.2018}, i.e., more difficult to identify even in very bright such as GX~5--1. \integral also detected a transient hard tail emission above 20--30 keV with the source in the HB \citep{Paizis.etAl.2006,Fabiani.etAl.2024}.

\subsection{Sco~X-1}

Sco~X-1 was the first extrasolar X-ray source discovered \citep{Giacconi.etAl.1962} and is the brightest persistent X-ray source in the sky. It is located at a distance of $2.13^{+0.21}_{-0.26}$ kpc, as derived from Gaia Data Release 2 \citep{Arnason.etAl.2021}. Its binary orbit has a low inclination angle to the observer \citep{Titarchuk.etAl.2014}, with an orbital period of 0.79 days. Sco~X-1 is a prototype of NS-LMXBs. It was classified as a Z-source based on its evolution along the CCD \citep{Hasinger.VanDerKlis.1989}, with a strong FB and large increases in X-ray intensity, distinct from Cyg~X-2-like sources, and with a barely visible HB. Sco~X-1 was the first X-ray binary system in which radio emission was observed \citep{Andrew.Purton.1968}. Very-long-baseline interferometry observations spatially resolved a radio jet at sub-milliarcsecond scales, characterized by the mildly relativistic motion of two components in opposite directions, with a position angle of 54$\degr$ and an inclination of 44$\degr$ \citep{Fomalont.etAl.2001,Fomalont.etAl.2001.b}.

Typical X-ray spectra of Sco~X-1 are characterized by soft thermal emission from the NS surface or the accretion disk and the Comptonization of these soft photons. The reflection component was observed in some spectra of Sco~X-1, exhibiting the Compton hump above 10 keV \citep{DiSalvo.etAl.2006}, as well as discrete features related to fluorescent emission and photoelectric absorption by heavy ions in the accretion disk. In particular, in addition to the ``classical'' Fe K$\alpha$ line at approximately 6.6 keV, the Fe K$\beta$ was also detected at 7.8 keV using \nustar data \citep{Mazzola.etAl.2021} and appears to be weaker along the FB. 

Sco~X-1 was also one of the sources observed during the first X-ray polarimetric campaign with the 8th Orbiting Solar Observatory (OSO-8). At 2.6 keV, the observed polarization degree (PD) was 0.4\% $\pm$ 0.2\% with a polarization angle (PA) of $29\degr \pm 10\degr$, while at 5.2 keV, OSO-8 measured a PD of 1.3\% $\pm$ 0.4\% with a PA of $57\degr \pm 6\degr$ \citep{Long.etAl.1979}. Before \ixpe, Sco~X-1 was observed with PolarLight \citep{Long.etAl.2022}. A significant detection was found only at higher energy during high-flux intervals, with a PD of 4.3\% $\pm$ 0.8\% and a PA of $53\degr \pm 5\degr$ in the 4--8 keV band. The PA observed by both satellites appears to be aligned with the direction of the radio jet. However, these results were obtained by integrating over the long exposure of the observations; therefore, the precise spectral state of the source could not be determined. 

\renewcommand{\arraystretch}{1.1}
\begin{table}
\caption{Log of the observations.}             
\label{table:Obs}      
\centering                                     
\begin{tabular}{l l c c}         
\hline\hline       
\noalign{\smallskip}
  Satellite & Obs. ID & Start Date [UTC] & Exp. [ks] \\  
\hline     
\noalign{\smallskip}
\multicolumn{4}{c}{Cyg~X-2} \\
  \ixpe & 01001601 & 2022-04-30 10:33:43 & 93.2 \\
  \ixpe & 01006601 & 2022-05-02 11:41:40 & 43.2 \\
  \nicer & 5034150102 & 2022-04-30 02:07:20 & 3.6 \\
  \nicer & 5034150103 & 2022-05-01 01:13:29 & 4.5 \\
  \nustar & 30801012002 & 2022-05-01 14:46:09 & 15.2 \\
  \hline
\multicolumn{4}{c}{XTE~J1071--461 (Obs. I)} \\
  \ixpe & 01250601 & 2022-09-29 12:46:08 & 47.7 \\
  \nicer & 5203390122 & 2022-09-29 00:51:09 & 0.5 \\
  \hline
\multicolumn{4}{c}{XTE~J1071--461 (Obs. II)} \\
  \ixpe & 01250701 & 2022-10-08 12:03:27 & 48.1 \\
  \nustar & 90801325002 & 2022-10-08 10:41:09 & 12.4 \\
  \hline
\multicolumn{4}{c}{GX~5--1 (Obs. I)} \\
  \ixpe & 02002701 & 2023-03-21 04:16:14 & 48.6 \\
  \nicer & 6010230101 & 2023-03-21 03:41:20 & 9.1 \\
  \nicer & 6010230102 & 2023-03-22 00:58:08 & 3.9 \\
  \nustar & 90902310002 & 2023-03-21 16:41:48 & 12.6 \\
  \hline
\multicolumn{4}{c}{GX~5--1 (Obs. II)} \\
  \ixpe & 02002702 & 2023-04-13 23:43:42 & 47.1 \\
  \nicer & 6010230106 & 2023-04-14 00:24:46 & 8.2 \\
  \nustar & 90902310004 & 2023-04-13 15:57:07 & 9.3 \\
  \nustar & 90902310006 & 2023-04-14 15:51:09 & 6.4 \\
  \hline
\multicolumn{4}{c}{Sco~X-1} \\
  \ixpe & 02002401 & 2023-08-28 16:10:56 & 24.2 \\
  \nicer & 6689010101 & 2023-08-28 15:23:00 & 3.0 \\
  \nicer & 6689010102 & 2023-08-29 00:40:40 & 2.5 \\
  \nustar & 30902036002 & 2023-08-28 17:41:09 & 1.0 \\
  \hline
\multicolumn{4}{c}{GX~340+0 (Obs. I)} \\
  \ixpe & 03003301 & 2024-03-23 13:56:25 & 99.0 \\
  \nustar & 91002313002 & 2024-03-27 06:16:09 & 12.5 \\
  \nustar & 91002313004 & 2024-03-28 11:11:09 & 12.5 \\
  \hline
\multicolumn{4}{c}{GX~340+0 (Obs. II)} \\
  \ixpe & 03009901 & 2024-08-12 07:25:09 & 191.6 \\
  \nicer & 7705010101 & 2024-08-12 06:18:20 & 3.1 \\
  \nicer & 7705010102 & 2024-08-13 00:52:20 & 4.1 \\
  \nicer & 7705010103 & 2024-08-14 00:06:07 & 4.0 \\
  \nicer & 7705010104 & 2024-08-15 02:24:28 & 2.5 \\
  \nicer & 7705010105 & 2024-08-16 00:04:43 & 1.4 \\
  \nustar & 30901012002 & 2024-08-17 23:16:09 & 22.5 \\
  \hline
\multicolumn{4}{c}{GX~349+2} \\
  \ixpe & 03003601 & 2024-09-06 22:52:51 & 95.6 \\
  \nustar & 91002333002 & 2024-09-07 05:51:09 & 7.7 \\
  \nustar & 91002333004 & 2024-09-08 12:16:09 & 10.1 \\
\hline      
\end{tabular}
\end{table}
\begin{figure*}[ht]
    \centering
    \begin{subfigure}[b]{0.325\textwidth}
    \centering
    \caption{Cyg~X-2}\label{fig:LC.CygX2}
    \includegraphics[width=\textwidth]{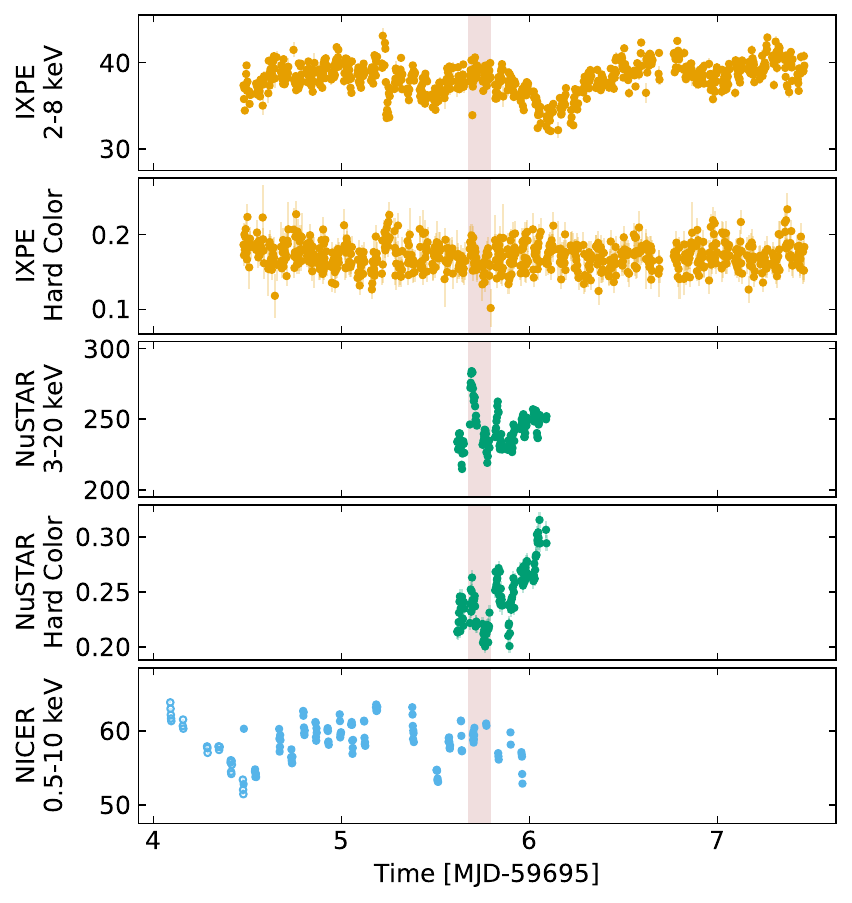}
    \end{subfigure}
    \hfill
    \begin{subfigure}[b]{0.325\textwidth}
    \centering
    \caption{XTE~J1701--462}\label{fig:LC.XTEJ1701}
    \includegraphics[width=\textwidth]{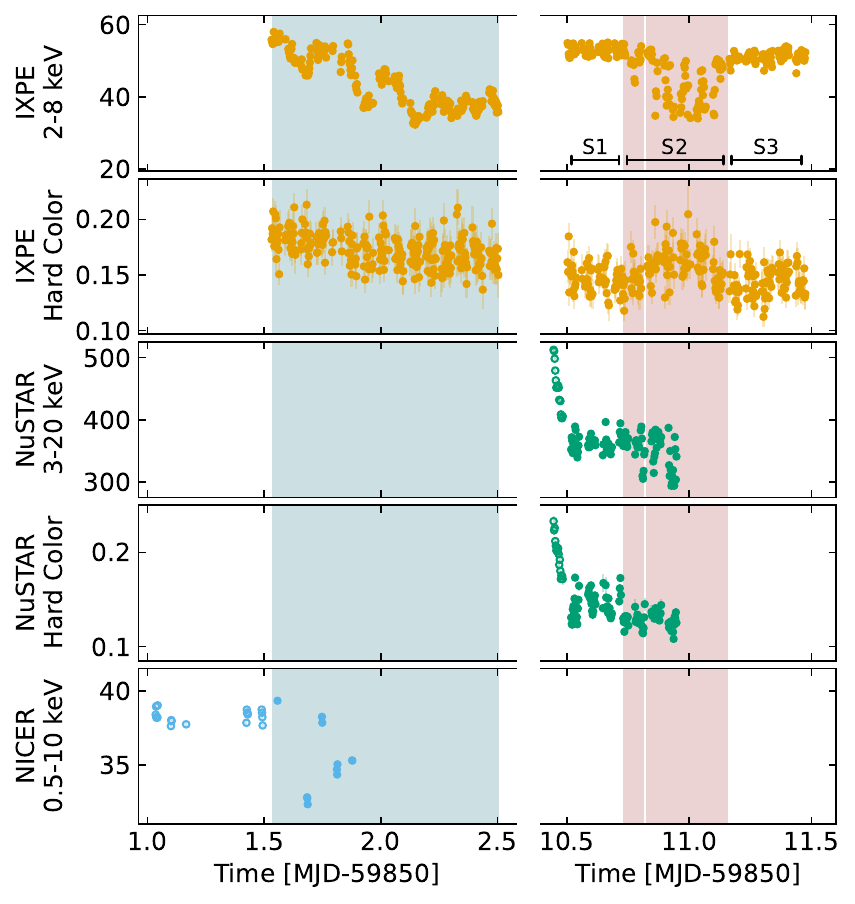}
    \end{subfigure}
    \hfill
    \begin{subfigure}[b]{0.325\textwidth}
    \centering
    \caption{GX~5--1}\label{fig:LC.GX5-1}
    \includegraphics[width=\textwidth]{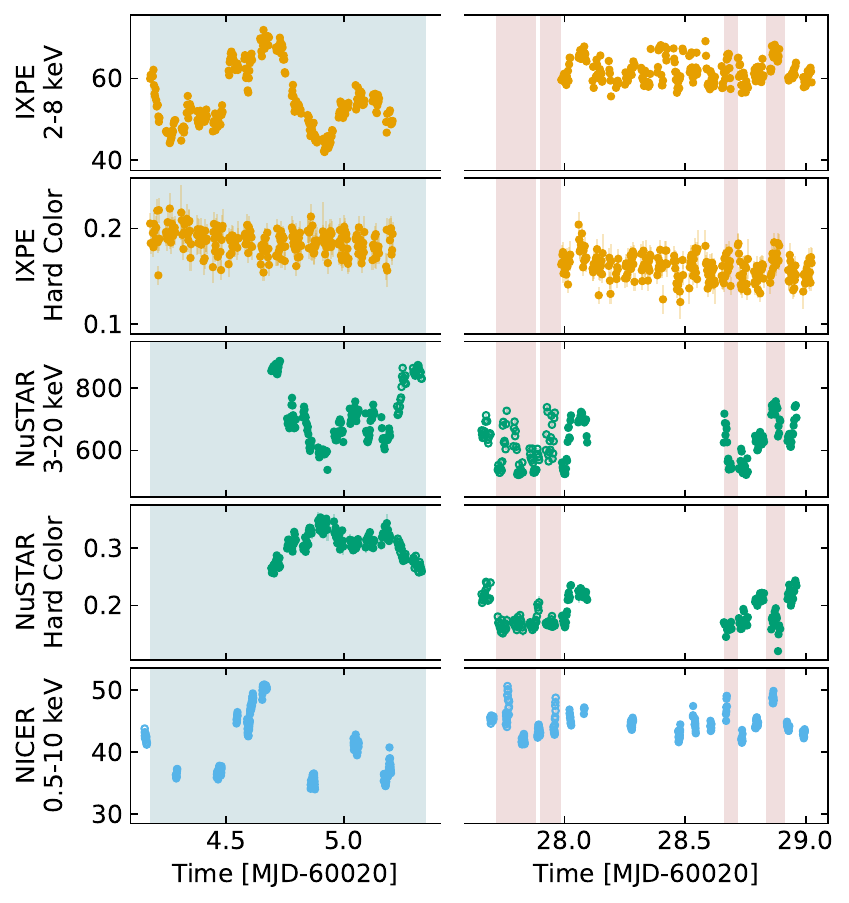}
    \end{subfigure}\\[1ex]
    \begin{subfigure}[b]{0.325\textwidth}
    \centering
    \caption{Sco~X-1}\label{fig:LC.ScoX-1}
    \includegraphics[width=\textwidth]{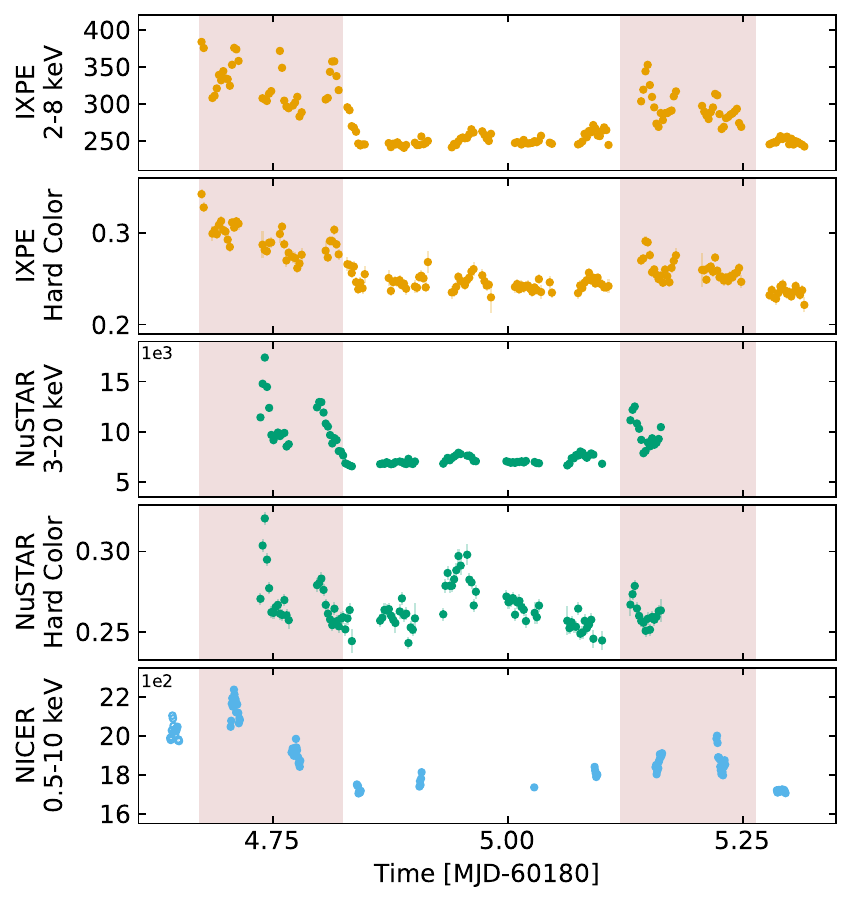}
    \end{subfigure}
    \hfill
    \begin{subfigure}[b]{0.325\textwidth}
    \centering
    \caption{GX~340+0}\label{fig:LC.GX340+0}
    \includegraphics[width=\textwidth]{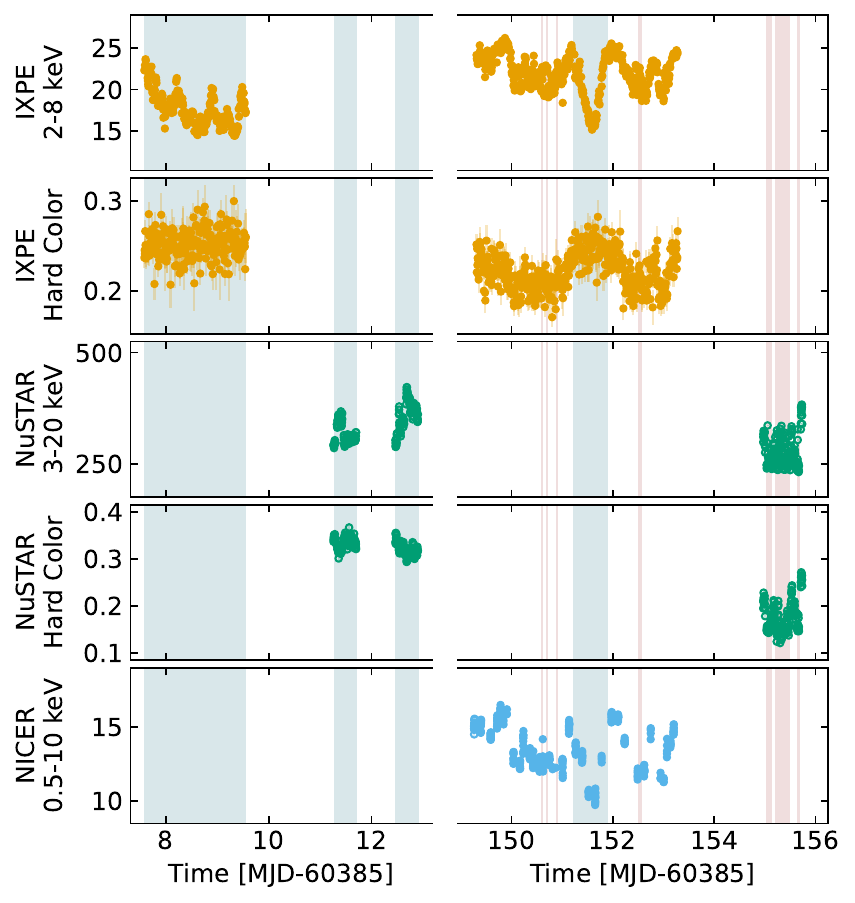}
    \end{subfigure}
    \hfill
    \begin{subfigure}[b]{0.325\textwidth}
    \centering
    \caption{GX~349+2}\label{fig:LC.GX349+2}
    \includegraphics[width=\textwidth]{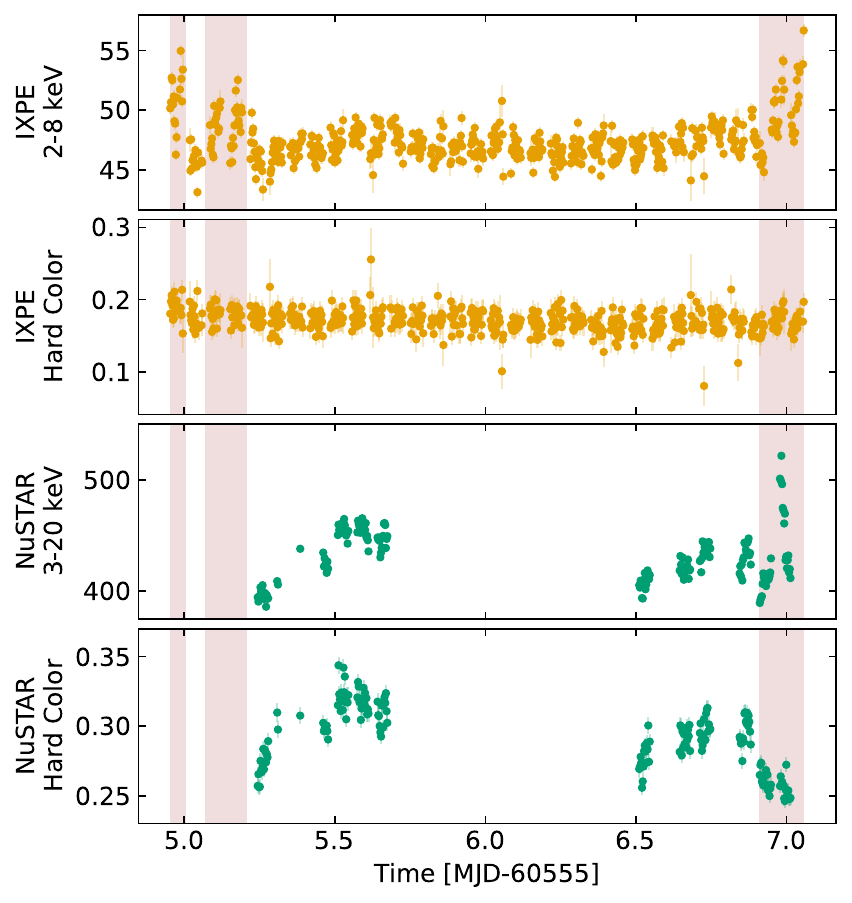}
    \end{subfigure}
    \caption{\ixpe, \nustar, and \nicer light curves (count\,s$^{-1}$) of each source. The second and fourth rows in each panel show the \ixpe hard color (5--8 keV/3--5 keV) and the \nustar hard color (10--20 keV/6--10 keV), respectively. Open circles indicate data points not simultaneous with \ixpe observation. Blue and red regions correspond to the HB and FB, respectively (see also Sect. \ref{sec:Timing} and \ref{sec:Polarization} for more details). Time bins of 200 s were used for \ixpe and \nustar, and 50 s for \nicer.}
    \label{fig:LC}
\end{figure*}

\subsection{GX~340+0}

GX~340+0 is a bright Cyg~X-2-like Z-source discovered by \citep{Friedman.etAl.1967,Rappaport.etAl.1971} with an Aerobee rocket, with a highly variable radio counterpart, from which a distance of $11 \pm 3$ kpc has been derived \citep{Fender.Hendry.2000}. This radio emission appears to be correlated with the X-ray flux along its HB \citep{Penninx.etAl.1993,Oosterbroek.etAl.1994,Berendsen.etAl.2000}, but without a clearly resolved radio jet. GX~340+0 typically covers the entire Z-track in a few days \citep{Jonker.etAl.2000.GX340+0}, mainly in the HB or NB, with rapid excursions along the FB \citep{Jonker.etAl.2000.GX340+0,Seifina.etAl.2013}.

The broadband spectrum was decomposed with a soft thermal component and a Comptonized emission. In addition, the relativistic iron line and several reflection features have been detected in GX~340+0 spectra \citep{Ueda.etAl.2005,DAi.etAl.2009,Cackett.etAl.2010}. Modeling the full reflection component leads to an inclination of about $35\degr-40\degr$ \citep{DAi.etAl.2009}, consistent with the lack of dips in the light curves \citep{Frank.etAl.1987,Kuulkers.etAl.1996}. 

\subsection{GX~349+2}

GX~349+2 is a bright Sco~X-1-like Z-source with a unique feature: while all known Z-sources trace the full Z-track during their evolution in their CCDs, GX~349+2 has been observed to cover only the NB or FB within a single day \citep{Hasinger.VanDerKlis.1989,Kuulkers.VanDerKlis1998,DiSalvo.etAl.2001,Iaria.etAl.2004,Cackett.etAl.2009,Kashyap.etAl.2023}. The distance to this source is approximately 9 kpc \citep{Grimm.etAl.2002}. Typical X-ray spectra are well characterized by a two-component model with a multicolor disk blackbody and Comptonization in the hot electron plasma of the corona \citep{DiSalvo.etAl.2001,Agrawal.Sreekumar.2003,Iaria.etAl.2004,Iaria.etAl.2009,Coughenour.etAl.2018}. Using BeppoSAX data, a strong hard tail was observed to dominate the spectrum above 30 keV during non-flaring activity \citep{DiSalvo.etAl.2001}, while all temperatures (i.e., those of the disk, the Comptonization region, and the seed-photons) tend to increase during flaring. At different epochs, a strong and variable relativistically broadened Fe K$\alpha$ emission line has been observed \citep{Cackett.etAl.2008,Cackett.etAl.2009,Iaria.etAl.2009,Coughenour.etAl.2018}. The inferred inclination of the system is approximately $25\degr-35\degr$ (depending on the model; \citealt{Coughenour.etAl.2018}), while the inner disk radius tends to remain at an average radius of 17.5 $R_{\rm g}$, but is not well constrained in the FB \citep{Coughenour.etAl.2018}. 

\begin{figure*}[ht]
    \centering
    \begin{subfigure}[b]{0.325\textwidth}
    \centering
    \caption{Cyg~X-2}\label{fig:CCD.CygX-2}
    \includegraphics[width=\textwidth]{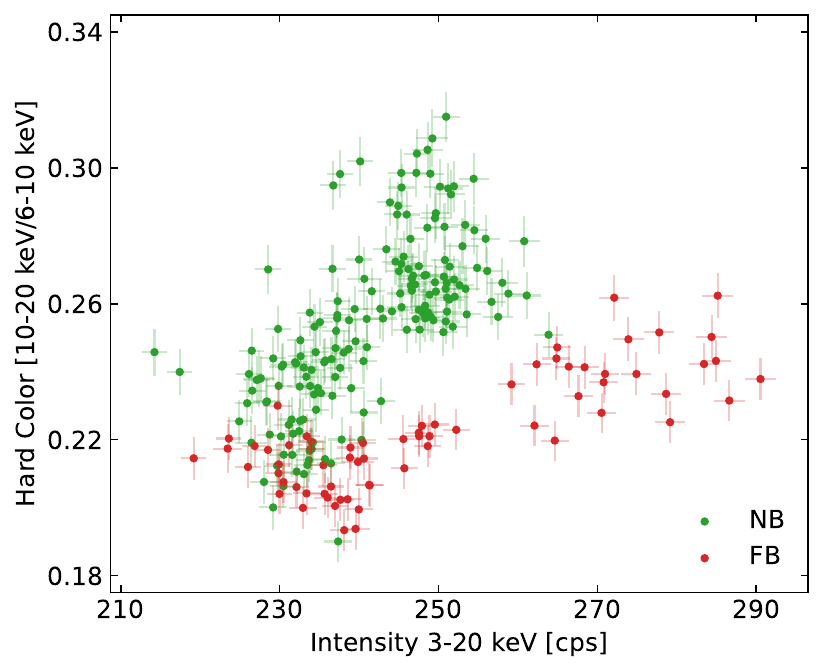}
    \end{subfigure}
    \begin{subfigure}[b]{0.325\textwidth}
    \centering
    \caption{XTE~J1701--462}\label{fig:CCD.XTEJ1701}    
    \includegraphics[width=\textwidth]{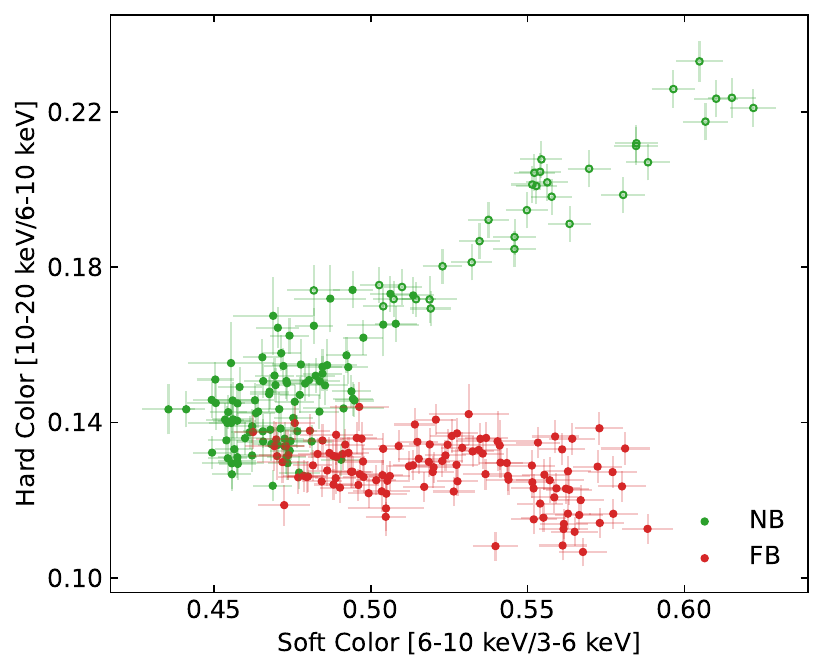}
    \end{subfigure}
    \begin{subfigure}[b]{0.325\textwidth}
    \centering
    \caption{GX~5--1}\label{fig:CCD.GX5-1}
    \includegraphics[width=\textwidth]{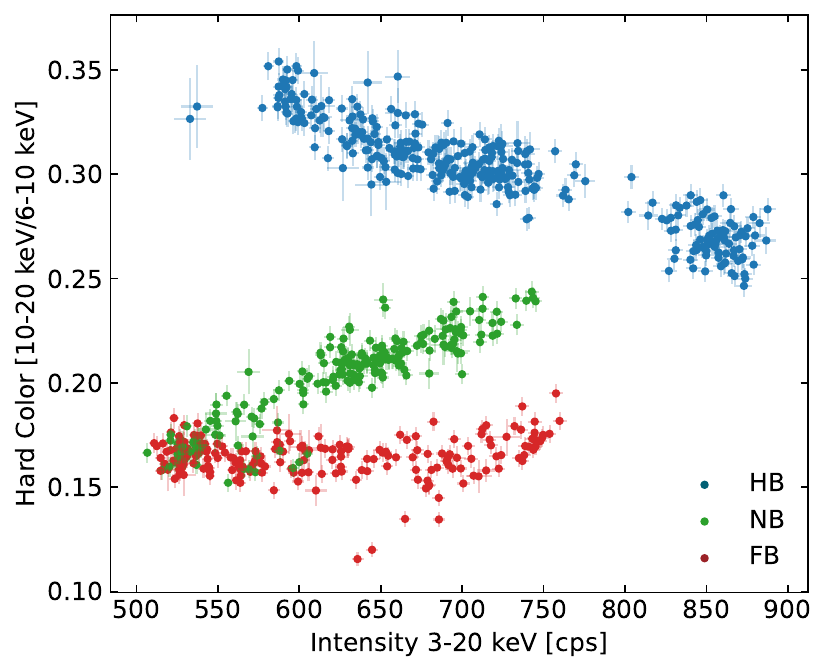}
    \end{subfigure}
    \\[1ex]
    \begin{subfigure}[b]{0.33\textwidth}
    \centering
    \caption{Sco~X-1}\label{fig:CCD.ScoX-1}
    \includegraphics[width=\textwidth]{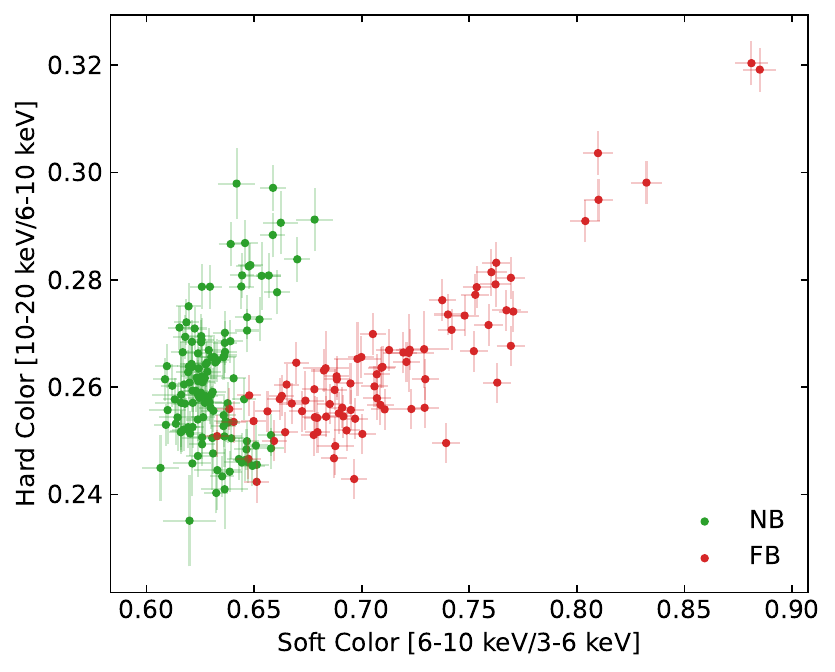}
    \end{subfigure}
    \begin{subfigure}[b]{0.325\textwidth}
    \centering
    \caption{GX~340+0}\label{fig:CCD.GX340+0}
    \includegraphics[width=\textwidth]{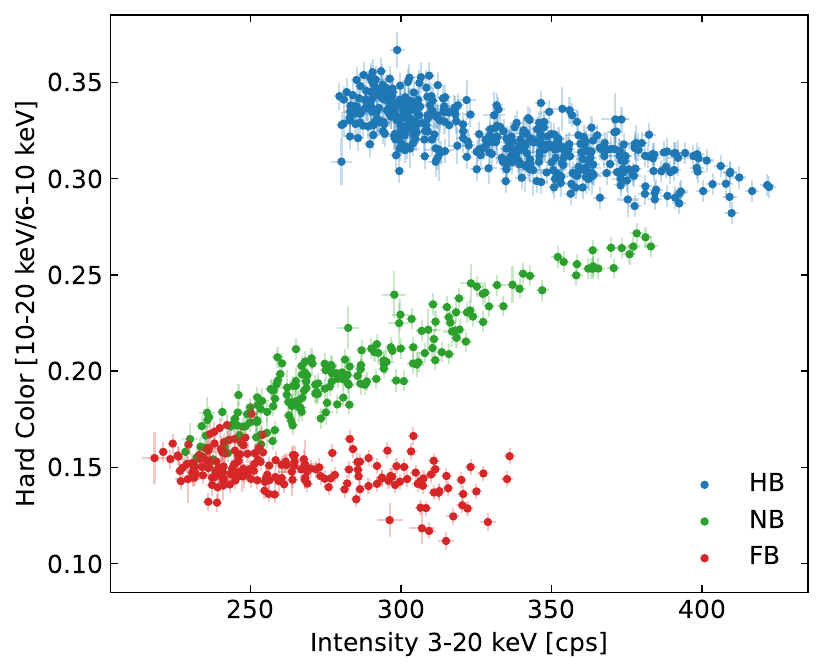}
    \end{subfigure}
    \begin{subfigure}[b]{0.325\textwidth}
    \centering
    \caption{GX~349+2}\label{fig:CCD.GX349+2}
    \includegraphics[width=\textwidth]{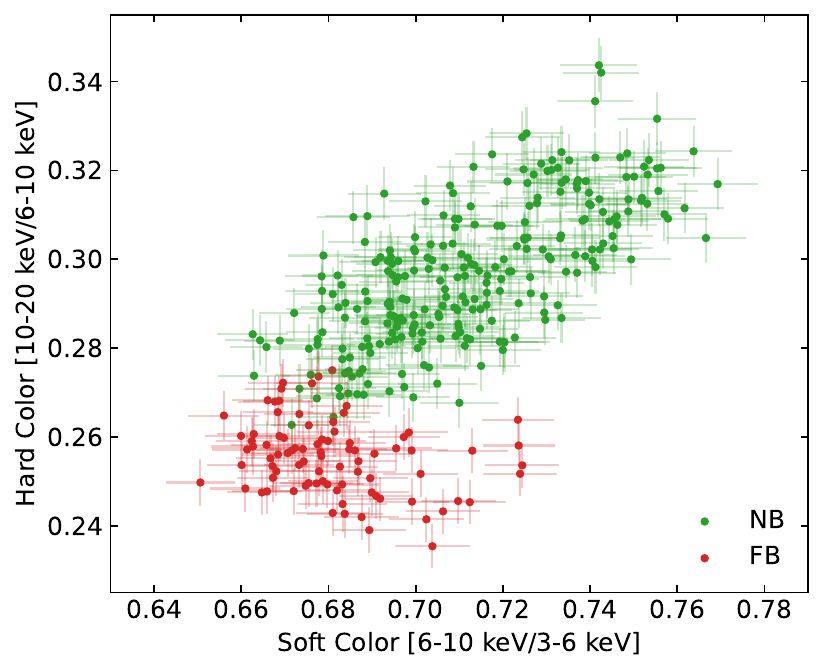}
    \end{subfigure}
    \caption{CCDs or HIDs using the \nustar observations (Table \ref{table:Obs}) of each source. The soft and hard colors are defined as the ratio of the counts in the 6--10 keV/3--6 keV and 10--20 keV/6--10 keV bands, respectively. The three branches are highlighted with different colors. We considered time bins of 200 s.}
    \label{fig:CCD}
\end{figure*}
\begin{figure*}[ht]
    \centering
    \begin{subfigure}[b]{0.3\textwidth}
    \centering
    \caption{Cyg~X-2}\label{fig:IXPE.CCD.CygX-2}
    \includegraphics[width=\textwidth]{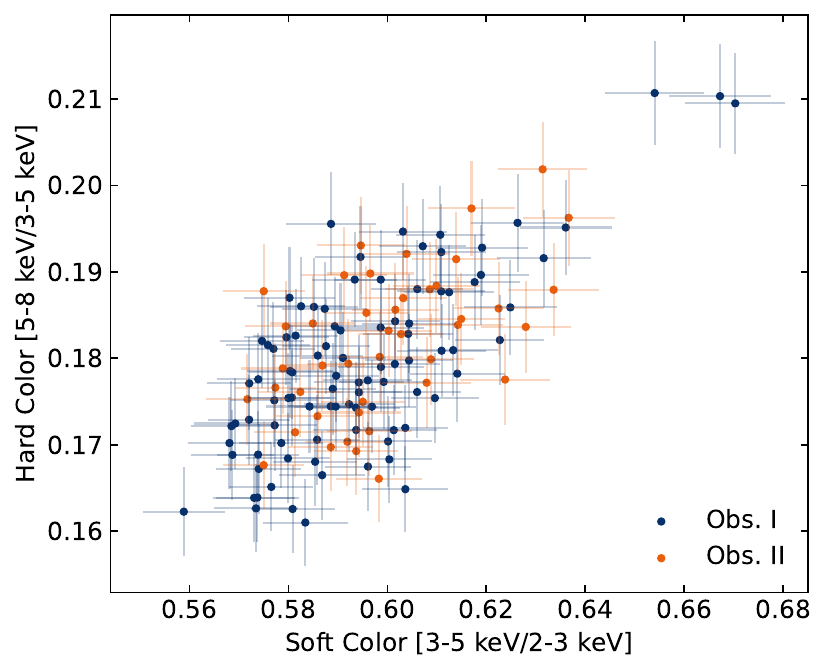}
    \end{subfigure}
    \begin{subfigure}[b]{0.3\textwidth}
    \centering
    \caption{XTE~J1701--462}\label{fig:IXPE.CCD.XTEJ1701}
    \includegraphics[width=\textwidth]{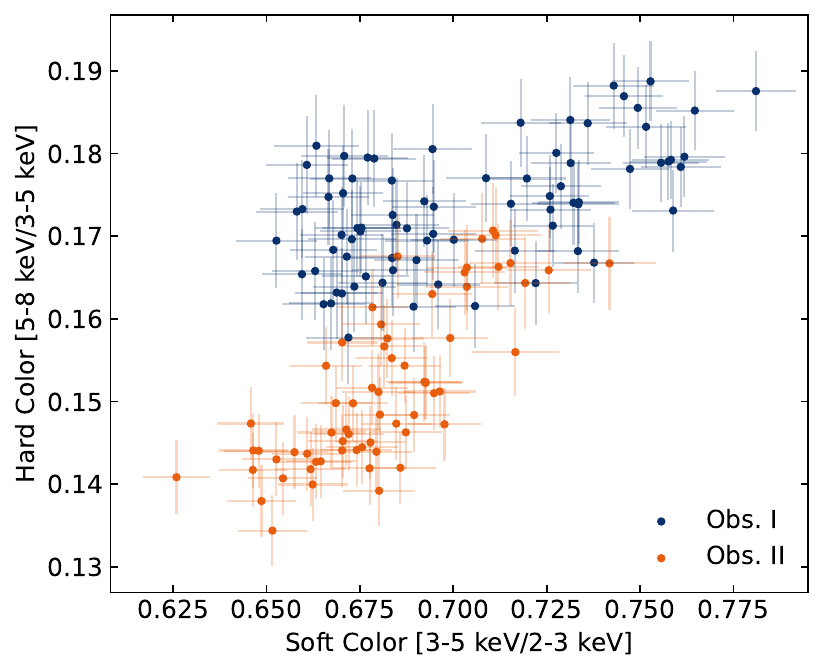}
    \end{subfigure}
    \begin{subfigure}[b]{0.3\textwidth}
    \centering
    \caption{GX~5--1}\label{fig:IXPE.CCD.GX5-1}
    \includegraphics[width=\textwidth]{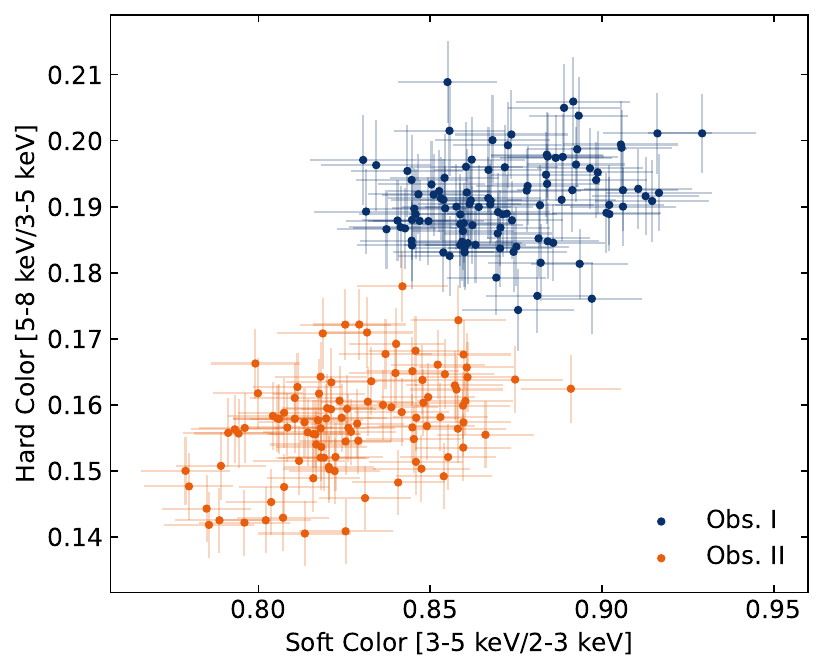}
    \end{subfigure}
    \\[1ex]
    \begin{subfigure}[b]{0.3\textwidth}
    \centering
    \caption{Sco~X-1}\label{fig:IXPE.CCD.ScoX-1}
    \includegraphics[width=\textwidth]{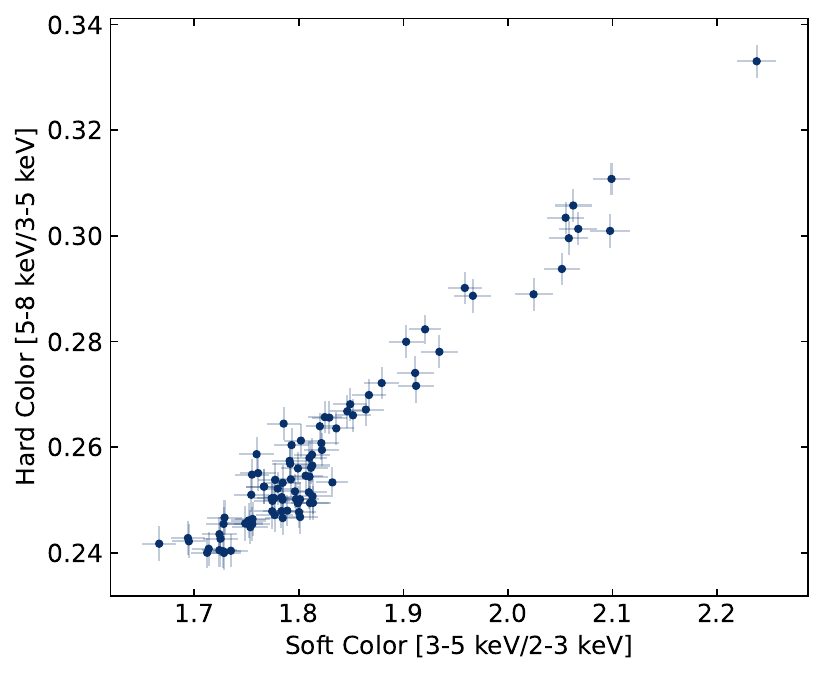}
    \end{subfigure}
    \begin{subfigure}[b]{0.3\textwidth}
    \centering
    \caption{GX~340+0}\label{fig:IXPE.CCD.GX340+0}
    \includegraphics[width=\textwidth]{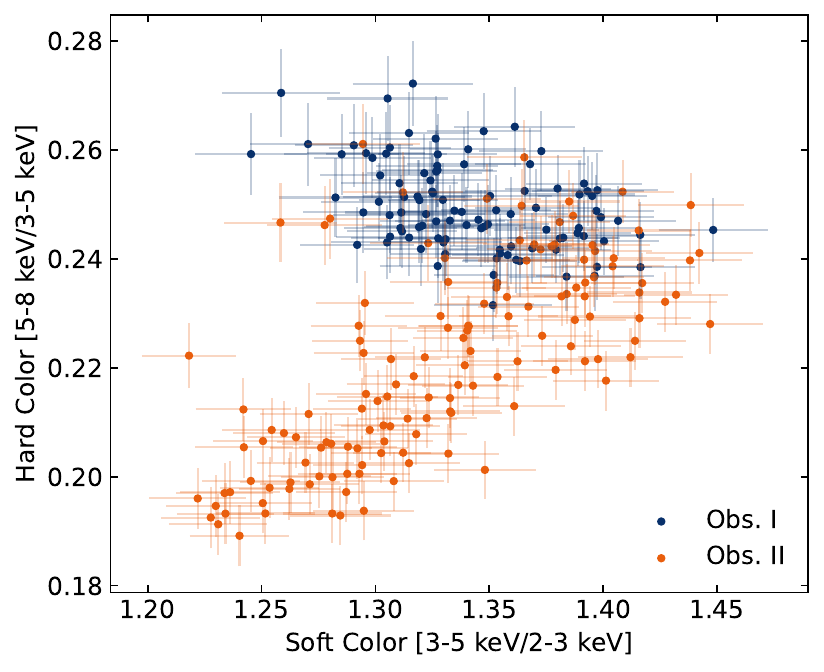}
    \end{subfigure}
    \begin{subfigure}[b]{0.3\textwidth}
    \centering
    \caption{GX~349+2}\label{fig:IXPE.CCD.GX349+2}
    \includegraphics[width=\textwidth]{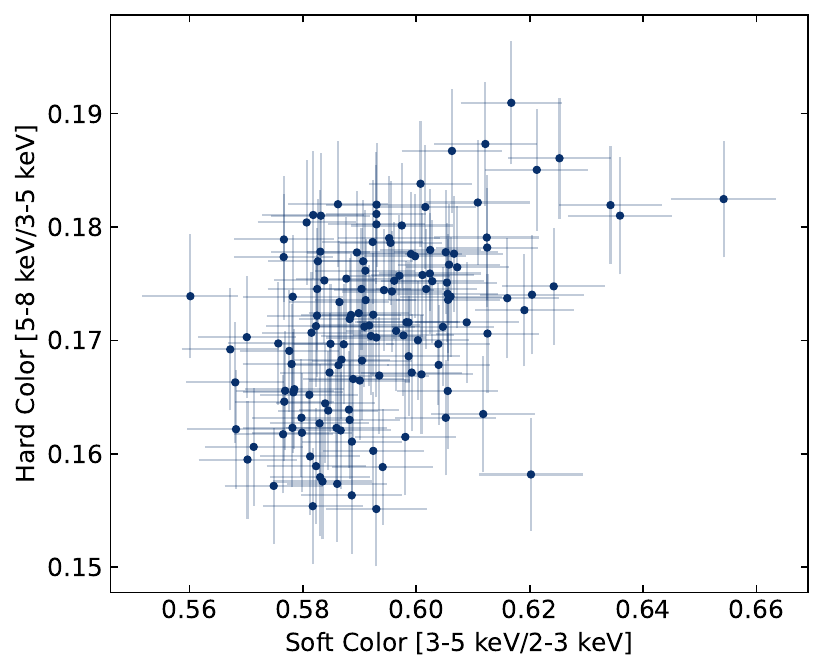}
    \end{subfigure}
    \caption{CCDs of each source using the \ixpe observations (Table \ref{table:Obs}). The soft and hard colors are defined as the ratio of the counts in the 3--5 keV/2--3 keV and 5--8 keV/3--5 keV bands, respectively. Different colors correspond to different observation of the source. Each bin corresponds to 1000 s.}
    \label{fig:IXPE.CCD}
\end{figure*}
\begin{figure*}[ht]
    \centering
    \begin{subfigure}[b]{0.2975\textwidth}
    \centering
    \caption{Cyg~X-2}\label{fig:Stokes.CygX-2}
    \includegraphics[width=\textwidth]{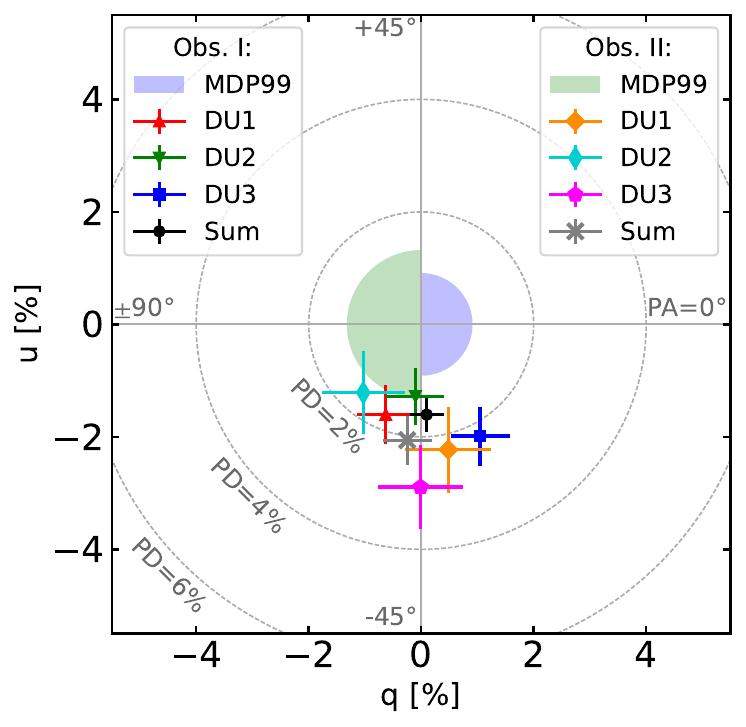}
    \end{subfigure}
    \begin{subfigure}[b]{0.2975\textwidth}
    \centering
    \caption{XTE~J1701--462}\label{fig:Stokes.XTEJ1701}
    \includegraphics[width=\textwidth]{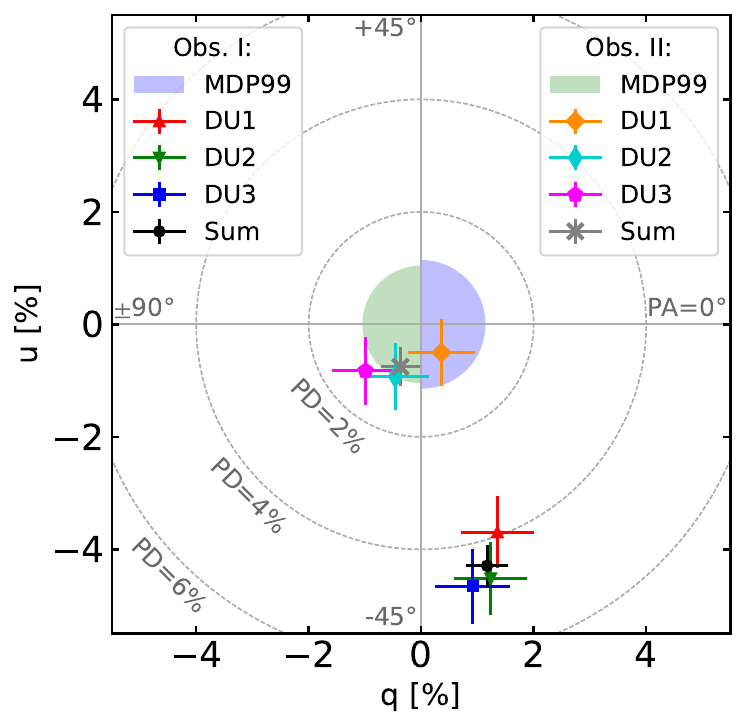}
    \end{subfigure}
    \begin{subfigure}[b]{0.2975\textwidth}
    \centering
    \caption{GX~5--1}\label{fig:Stokes.GX5-1}
    \includegraphics[width=\textwidth]{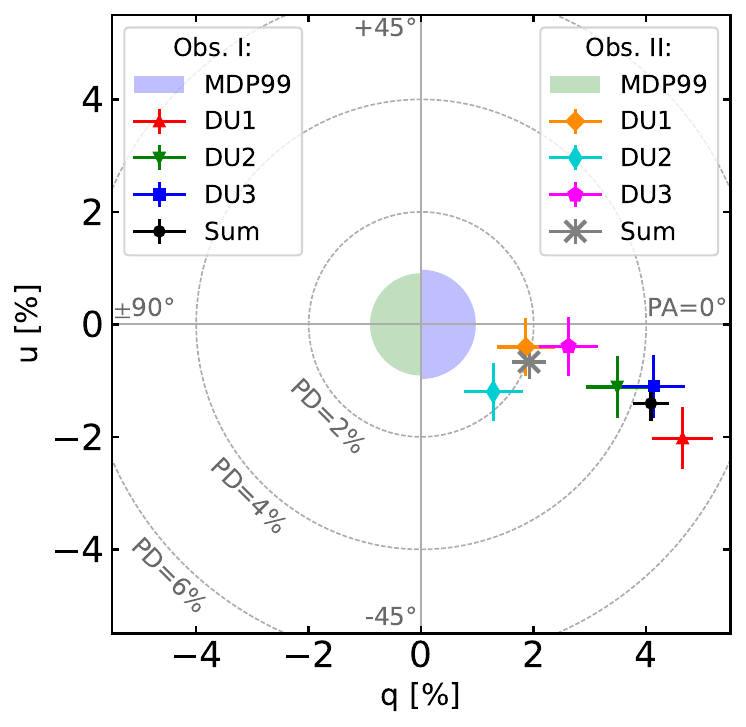}
    \end{subfigure}
    \\[1ex]
    \begin{subfigure}[b]{0.2975\textwidth}
    \centering
    \caption{Sco~X-1}\label{fig:Stokes.ScoX-1}
    \includegraphics[width=\textwidth]{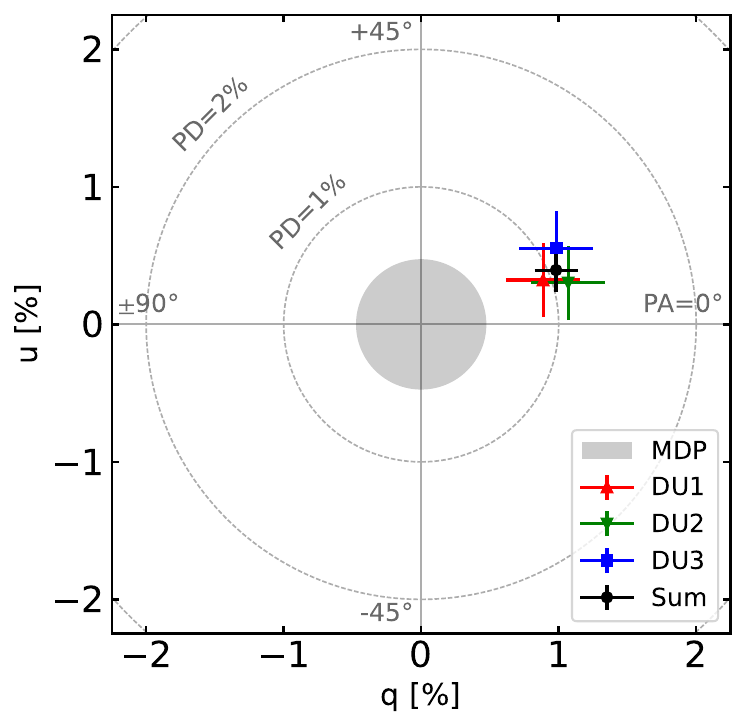}
    \end{subfigure}
    \begin{subfigure}[b]{0.2975\textwidth}
    \centering
    \caption{GX~340+0}\label{fig:Stokes.GX340+0}
    \includegraphics[width=\textwidth]{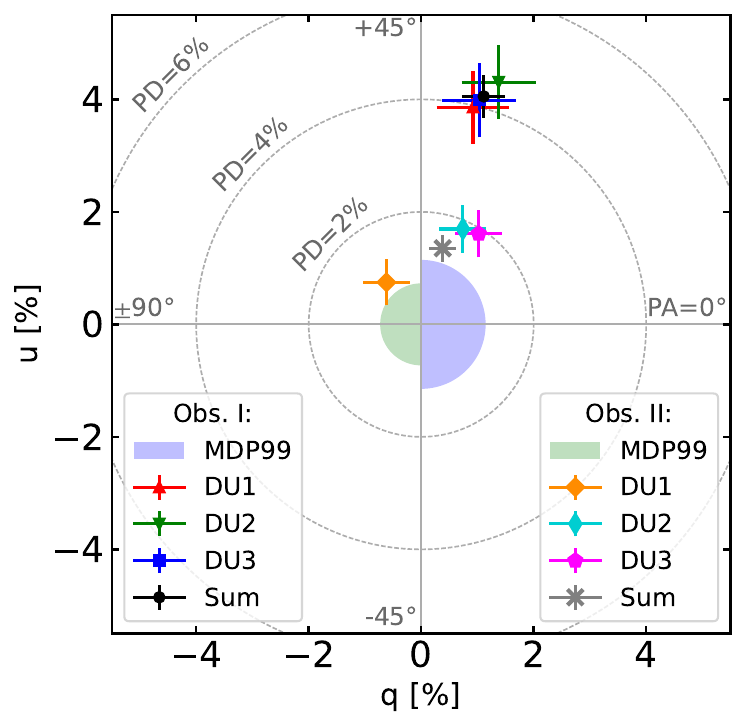}
    \end{subfigure}
    \begin{subfigure}[b]{0.2975\textwidth}
    \centering
    \caption{GX~349+2}\label{fig:Stokes.GX349+2}
    \includegraphics[width=\textwidth]{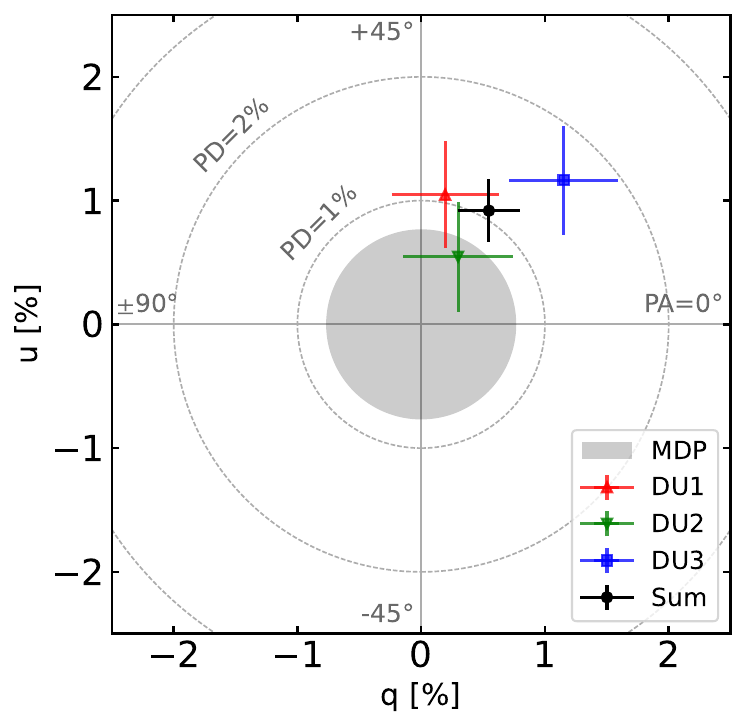}
    \end{subfigure}
    \caption{Normalized Stokes parameters $q$ ($Q/I$) and $u$ ($U/I$) in the 2--8 keV energy range, integrated over the entire \ixpe observations obtained with \texttt{PCUBE} \citep{Baldini.etAl.2022}, for the three \ixpe DUs and their sum. The filled regions corresponds to the minimum detectable polarization at the 99\% confidence level (MDP99).}
    \label{fig:Stokes}
\end{figure*}


\section{Observations and data reduction}\label{sec:Observations}

\subsection{\ixpe}

All \ixpe observations of Z-sources are listed in Table \ref{table:Obs}. The time-resolved polarimetric data for each detector unit (DU) were produced using \textsc{ixpeobssim} \citep[v.31.0.3][]{Baldini.etAl.2022} with the latest calibration files (v.20240701.13). The Stokes parameters, PD, PA, and the associated uncertainties were calculated with the unweighted, model-independent \texttt{PCUBE} binning algorithm of \textsc{ixpeobssim}. For data extraction, we considered a circular region centered on the source, while the background was extracted from an annular region centered on the source with an inner and outer radius of 180\arcsec\ and 240\arcsec, respectively. We obtained the radius of the extraction region via an iterative process that maximizes the signal-to-noise ratio (S/N) in the 2--8 keV band\footnote{For GX~5--1, we used a fixed $60\arcsec$ radius region to avoid the X-ray halo detected by \textit{Chandra} \citep{Smith.etAl.2006,Clark.2018}.}, similar to the procedure described in \cite{Piconcelli.etAl.2004}. As all of these sources are very bright ($>2$\,counts\,s$^{-1}$\,arcmin$^{-2}$), neither background rejection nor subtraction was performed, according to the prescription of \cite{DiMarco.etAl.2023}. Due to the extremely high X-ray flux of Sco~X-1, the \ixpe observation was performed with a partially opaque absorber (i.e., gray filter) mounted on the onboard calibration wheel \citep{Ferrazzoli.etAl.2020,Soffitta.etAl.2021,Weisskopf.2022} to reduce the flux to a level compatible with the dead time of each \ixpe DU.

Although the \ixpe energy band is limited, we also constructed the CCD for each observed source to characterize their state during the \ixpe observations. In particular, we defined the \ixpe soft and hard color as the ratios of the measured counts in the 3--5/2--3 keV and 5--8/3--5 keV energy bands, respectively.

\subsection{\nicer}

The Neutron Star Interior Composition Explorer (\nicer; \citealt{Gendreau.etAl.2016}) observed most of the sources simultaneously with \ixpe (see Table \ref{table:Obs}). We extracted the calibrated and cleaned files using the \texttt{nicerl2} task of the \nicer Data Analysis Software (\textsc{nicerdas} v.13), together with the latest calibration files (CALDB v.20240206). The light curves used in this part of the work were derived with the \texttt{nicerl3-lc} command. 

\subsection{\nustar}

The Nuclear Spectroscopic Telescope Array (\nustar; \citealt{Harrison.etAl.2013}) observed most of the Z-sources considered simultaneously with \ixpe (see Table \ref{table:Obs}). The unfiltered data were reduced using the latest calibration files (CALDB v.20240812) with the standard \texttt{nupipeline} task of the \nustar Data Analysis Software (\textsc{nustardas} v.2.1.4). Since all sources are particularly bright ($>100$\,counts\,s$^{-1}$), we used the keyword \texttt{statusexpr="(STATUS==b0000xxx00xxxx000)\&\&(SHIELD= =0)"} during the \texttt{nupipeline} task. For each focal plane module (FPM), we extracted the light curves using the \texttt{nuproducts} task, using a circular region centered on the sources. The extraction radii were derived using the same procedure to maximize the S / N as used for \ixpe data reduction (see also \citealt{Piconcelli.etAl.2004}). Since the background is not negligible at all energies, we performed background subtraction for the \nustar data. For both FPMs, we extracted the background from a circular region with 60\arcsec radius, sufficiently far from the source. For each source observed with \nustar, we constructed the CCD or HID to characterize the state of the source and study its evolution. We defined soft and hard color as the ratios of the counts in the 6--10/3--6 keV and 10--20/6--10 keV energy ranges, respectively. 


\section{Timing properties}\label{sec:Timing}

The light curves obtained with \ixpe, \nustar, and \nicer are shown in Fig. \ref{fig:LC}, using time bins of 200 s for \ixpe and \nustar, and 50 s for \nicer. Since Z-sources are variable objects and their polarization properties appear to be correlated with the position in the CCDs or HIDs, it is important to identify the position of each source along the Z-track. With the exception of the first observation of XTE~J1701--462 and those of GX~340+0, all other sources have \nustar observations simultaneous with \ixpe. Due to the limited band-pass of \ixpe and \nicer, it is difficult to distinguish between the three branches, particularly to identify the FB. \nustar enables the construction of proper CCDs and HIDs to characterize the state of the sources in detail. The \nustar CCDs or HIDs for each source are shown in Fig. \ref{fig:CCD}. The choice to display the CCD for some sources and the HID for others is purely for graphical clarity: for Z-sources with the HB detected, this branch is more distinctly visible in the HID than in the CCD, whereas for other sources, the FB is better identified using the CCD instead of the corresponding HID. All Z-sources were observed at least once in the NB and in the FB. The FB intervals are highlighted with red regions in Fig. \ref{fig:LC}, while the HB intervals correspond to blue regions in Fig. \ref{fig:LC}. Only three of the Z-sources considered (XTE~J1701--462, GX~5--1, and GX~340+0) were observed in the HB. In particular, GX~5--1 is the only one with both \nustar and \nicer observations simultaneous with \ixpe that exhibits the full Z-track (Fig. \ref{fig:LC.GX5-1}). XTE~J1701--462 was observed only by \nicer during the first \ixpe observation with the source in the HB (Fig. \ref{fig:IXPE.CCD.XTEJ1701}; see also \citealt{Cocchi.etAl.2023}), while during the second observation only \nustar observed the source tracking the NB and FB simultaneously with the \ixpe exposure. Moreover, we further divided the second observations into three segments, highlighted with black intervals in Fig. \ref{fig:LC.XTEJ1701}, similarly to \cite{Zhao.etAl.2024}. During the first and third intervals, while the source moves along the NB, the \ixpe count rate and hard color are quite similar, and they vary during the second interval when the source moves along the FB. GX~340+0 was not observed simultaneously with \ixpe by \nicer or \nustar during the first segment (Fig. \ref{fig:LC.GX340+0}). However, from \ixpe data (Fig. \ref{fig:IXPE.CCD.GX340+0}), the source can be identified in the HB. It remains in the same branch during the first two \nustar observations of this source, as indicated by its HID (Fig. \ref{fig:CCD.GX340+0}). During the second \ixpe segment, the source mainly moves along the NB with a brief excursion into the HB (Fig. \ref{fig:IXPE.CCD.GX340+0}),as indicated by the short black interval in Fig. \ref{fig:LC.GX340+0}. In contrast, the source tracks the NB and FB during the \nustar exposure, a few days after the \ixpe observation. \cite{Bhargava.etAl.2024b} also identified the FB using \textit{AstroSat} data simultaneous with the \ixpe observation; therefore, we considered the same time intervals (see Table 2 of \citealt{Bhargava.etAl.2024a}) to compute the polarization in the FB. For each source, we then created good time intervals (GTIs) for each detected branch and extracted all the data from each instrument to study the polarimetric properties of the Z-sources along the CCDs and HIDs in detail. 


\section{Model-independent polarimetric analysis}\label{sec:Polarization}

\begin{figure*}
    \centering
    \begin{subfigure}[b]{0.2975\textwidth}
    \centering
    \caption{Cyg~X-2}\label{fig:Contour.CygX-2}
    \vspace{-0.5ex}
    \includegraphics[width=\textwidth]{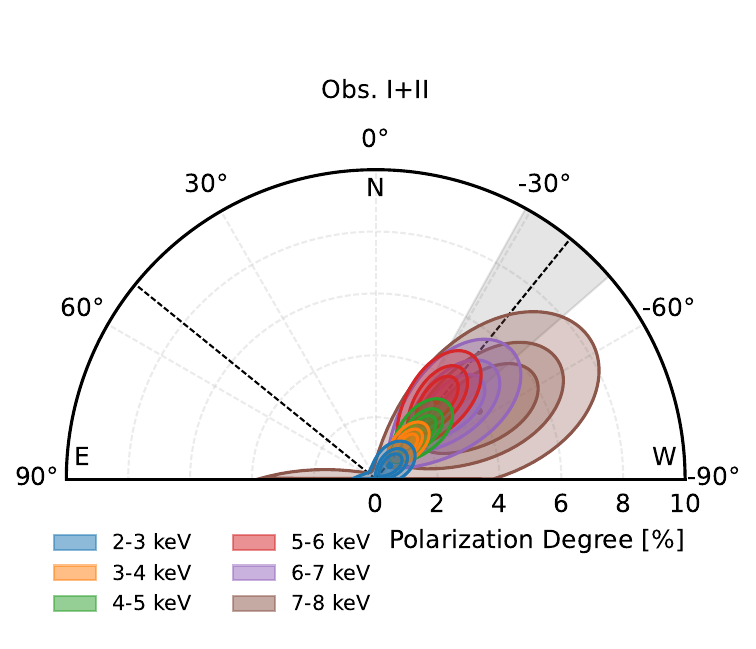}
    \end{subfigure}
    \begin{subfigure}[b]{0.2975\textwidth}
    \centering
    \caption{XTE~J1701--462 (Obs. I)}\label{fig:Contour.XTEJ1701.ObsI}
    \vspace{-0.5ex}
    \includegraphics[width=\textwidth]{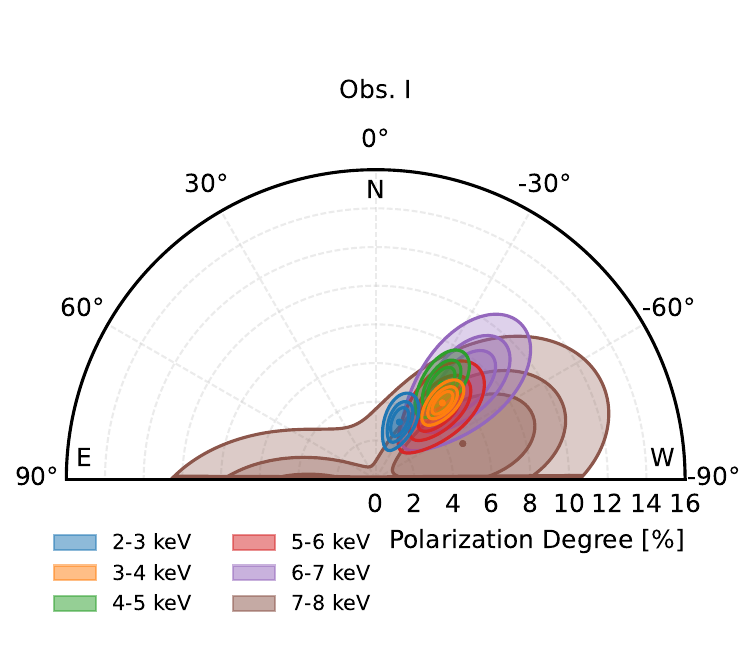}
    \end{subfigure}
    \begin{subfigure}[b]{0.2975\textwidth}
    \centering
    \caption{XTE~J1701--462 (Obs. II)}\label{fig:Contour.XTEJ1701.ObsII}
    \vspace{-0.5ex}
    \includegraphics[width=\textwidth]{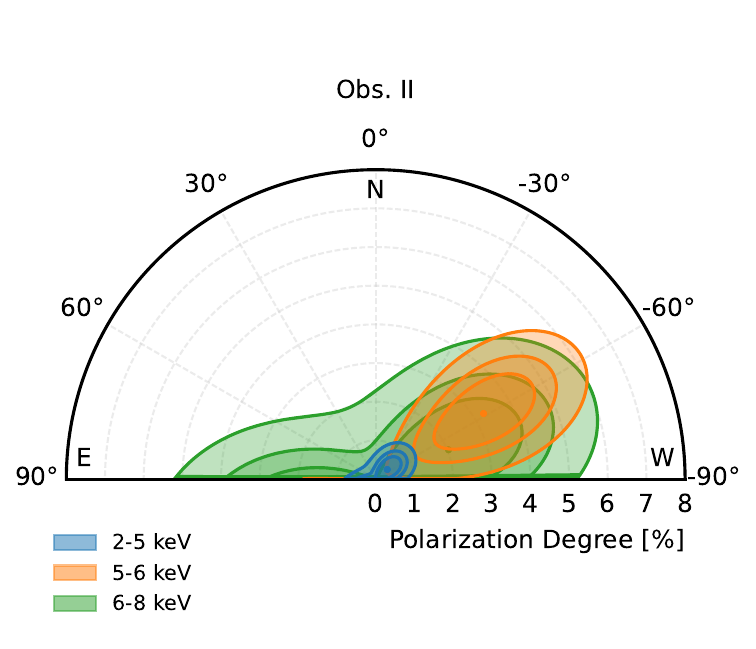}
    \end{subfigure}
    \\[1ex]
    \begin{subfigure}[b]{0.2975\textwidth}
    \centering
    \caption{GX~5--1 (Obs. I)}\label{fig:Contour.GX5-1.ObsI}
    \vspace{-0.5ex}
    \includegraphics[width=\textwidth]{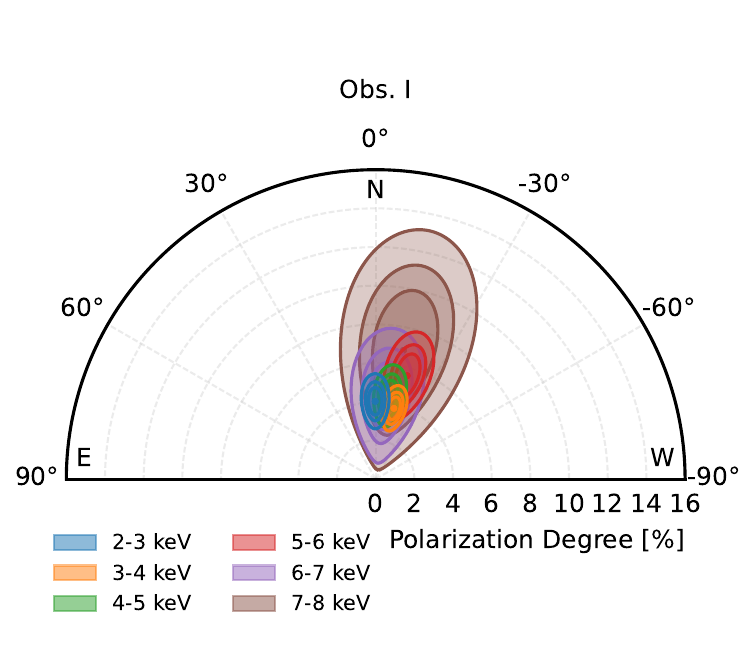}
    \end{subfigure}
    \begin{subfigure}[b]{0.2975\textwidth}
    \centering
    \caption{GX~5--1 (Obs. II)}\label{fig:Contour.GX5-1.ObsII}
    \vspace{-0.5ex}
    \includegraphics[width=\textwidth]{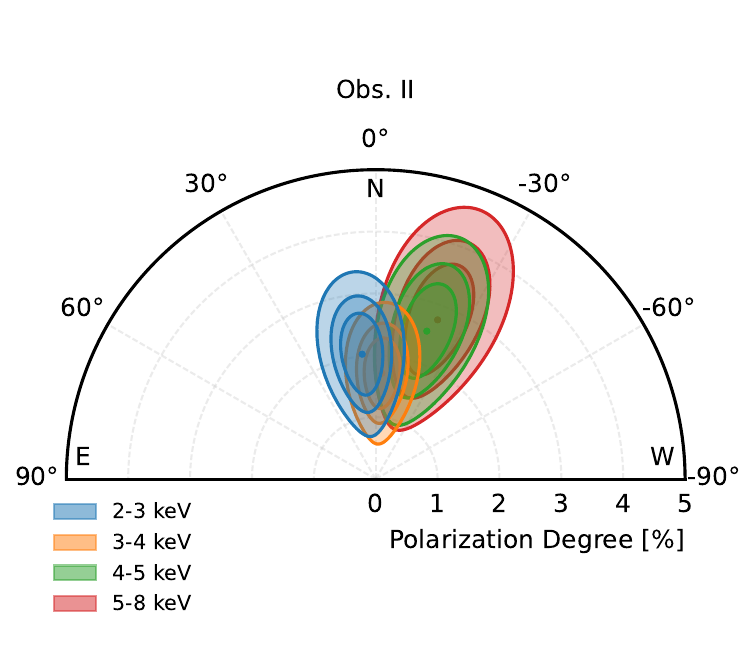}
    \end{subfigure}
    \begin{subfigure}[b]{0.2975\textwidth}
    \centering
    \caption{Sco~X-1}\label{fig:Contour.ScoX-1}
    \vspace{-0.5ex}
    \includegraphics[width=\textwidth]{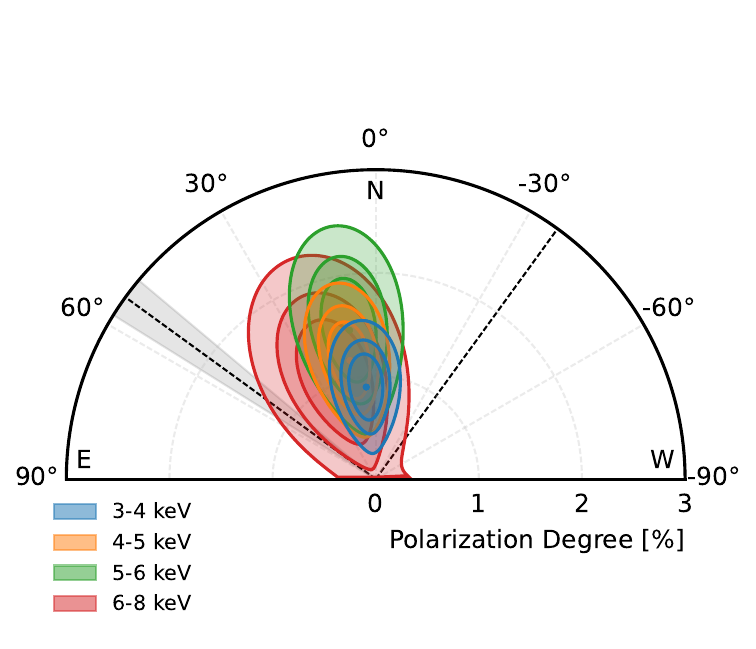}
    \end{subfigure}
    \\[1ex]
    \begin{subfigure}[b]{0.2975\textwidth}
    \centering
    \caption{GX~340+0 (Obs. I)}\label{fig:Contour.GX340+0.ObsI}
    \vspace{-0.5ex}
    \includegraphics[width=\textwidth]{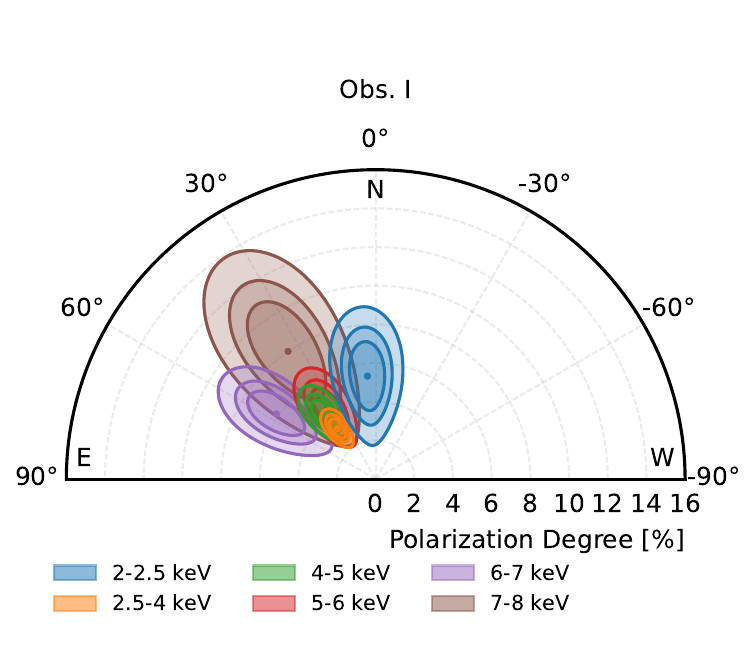}
    \end{subfigure}
    \begin{subfigure}[b]{0.2975\textwidth}
    \centering
    \caption{GX~340+0 (Obs. II)}\label{fig:Contour.GX340+0.ObsII}
    \vspace{-0.5ex}
    \includegraphics[width=\textwidth]{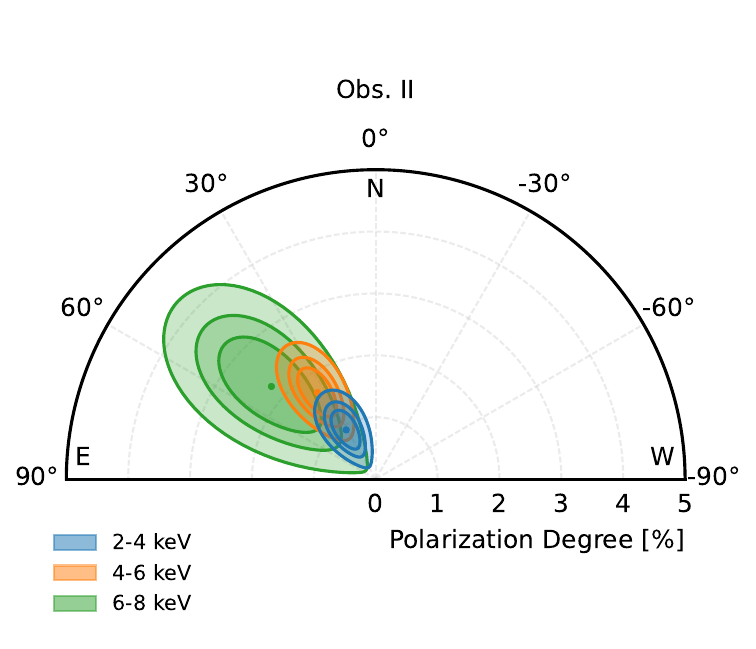}
    \end{subfigure}
    \begin{subfigure}[b]{0.2975\textwidth}
    \centering
    \caption{GX~349+2}\label{fig:Contour.GX349+2}
    \vspace{-0.5ex}
    \includegraphics[width=\textwidth]{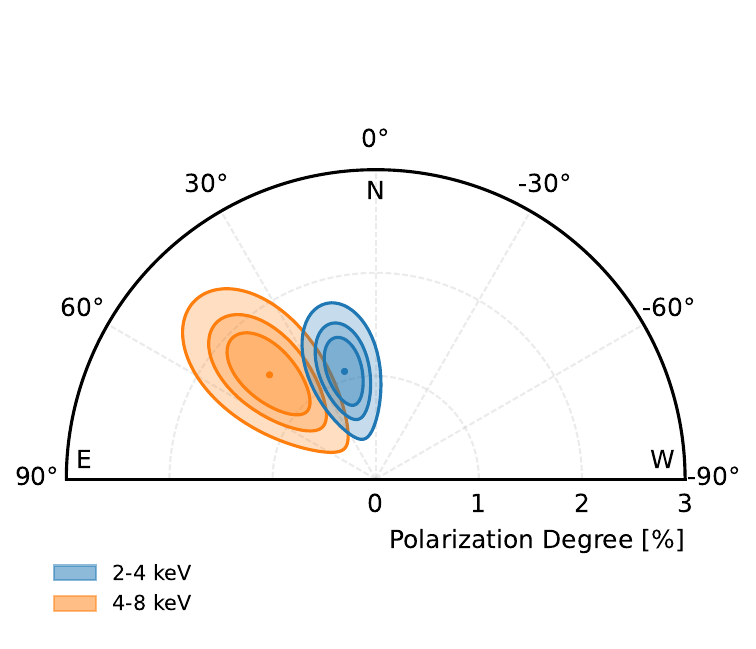}
    \end{subfigure}
    \caption{Polarization contours at the 68\%, 90\%, and 99\% confidence levels, computed using Eq. (32) of \cite{Muleri.2022}. Different colors correspond to different energy bins. For Cyg~X-2 and Sco~X-1, the gray region corresponds to the jet direction.}
    \label{fig:Contours}
\end{figure*}
\begin{figure*}
    \centering
    \begin{subfigure}[b]{0.2975\textwidth}
    \centering
    \caption{Cyg~X-2}\label{fig:PD-En.CygX-2}
    \includegraphics[width=\textwidth]{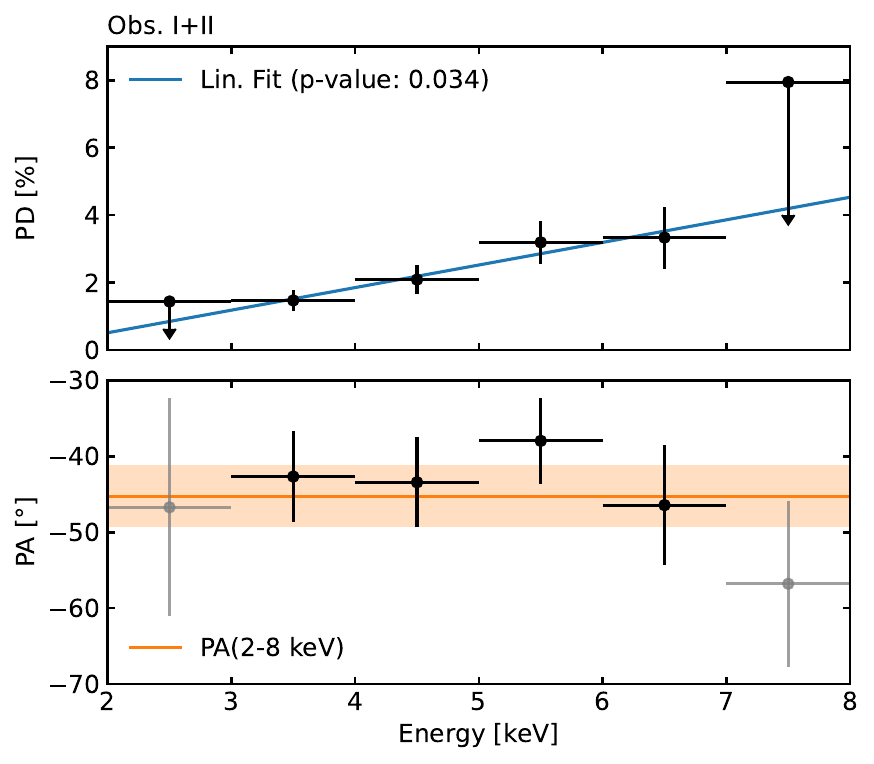}
    \end{subfigure}
    \begin{subfigure}[b]{0.2975\textwidth}
    \centering
    \caption{XTE~J1701--462 (Obs. I)}\label{fig:PD-En.XTEJ1701.ObsI}
    \includegraphics[width=\textwidth]{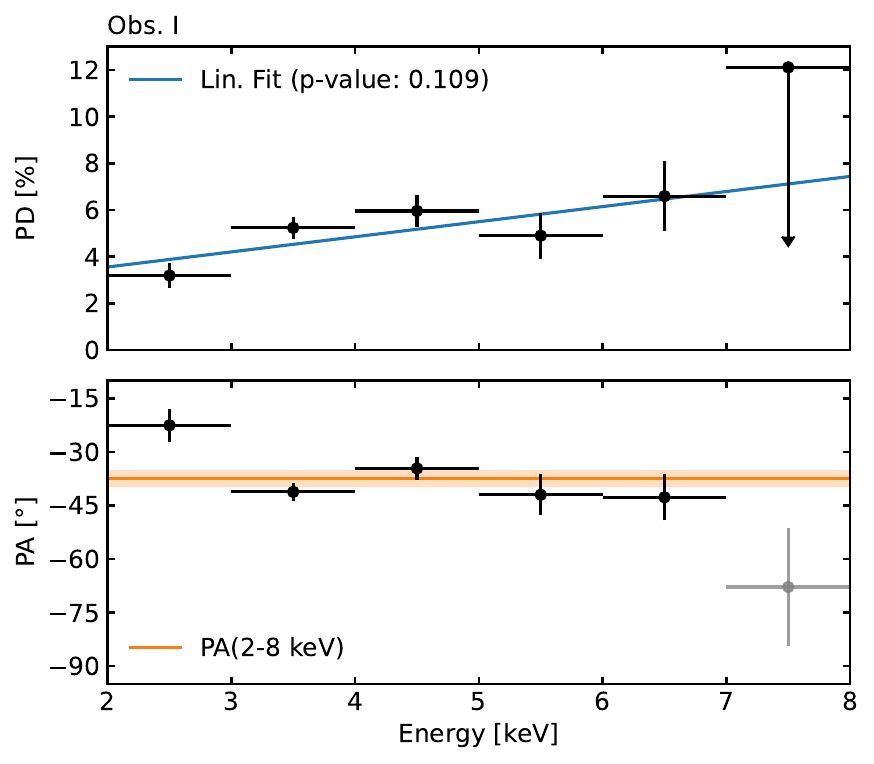}
    \end{subfigure}
    \begin{subfigure}[b]{0.2975\textwidth}
    \centering
    \caption{XTE~J1701--462 (Obs. II)}\label{fig:PD-En.XTEJ1701.ObsII}
    \includegraphics[width=\textwidth]{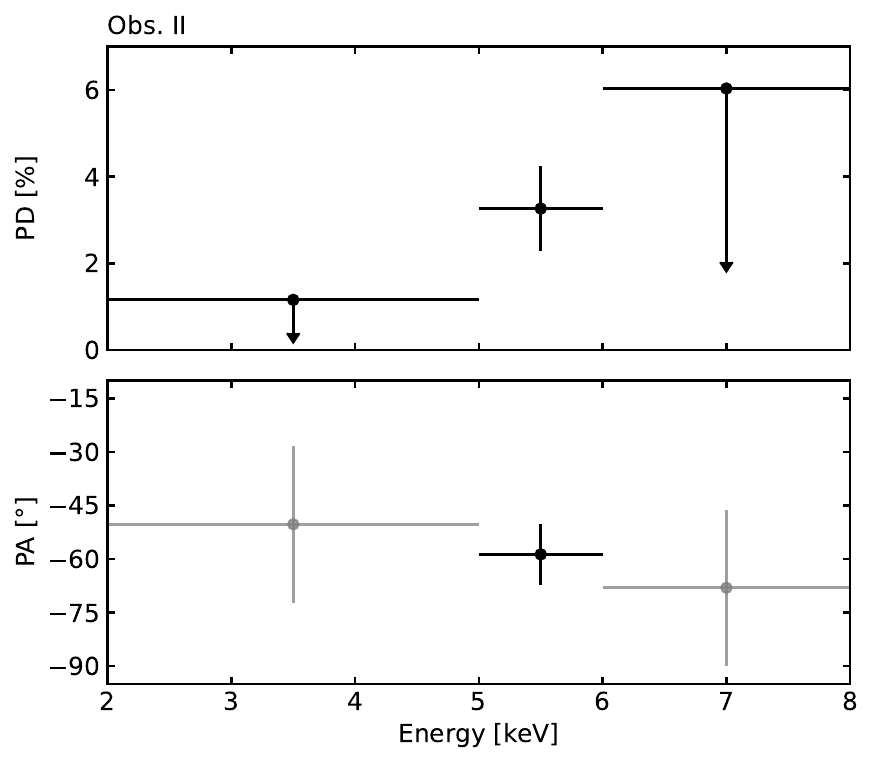}
    \end{subfigure}
    \\[1ex]
    \begin{subfigure}[b]{0.2975\textwidth}
    \centering
    \caption{GX~5--1 (Obs. I)}\label{fig:PD-En.GX5-1.ObsI}
    \includegraphics[width=\textwidth]{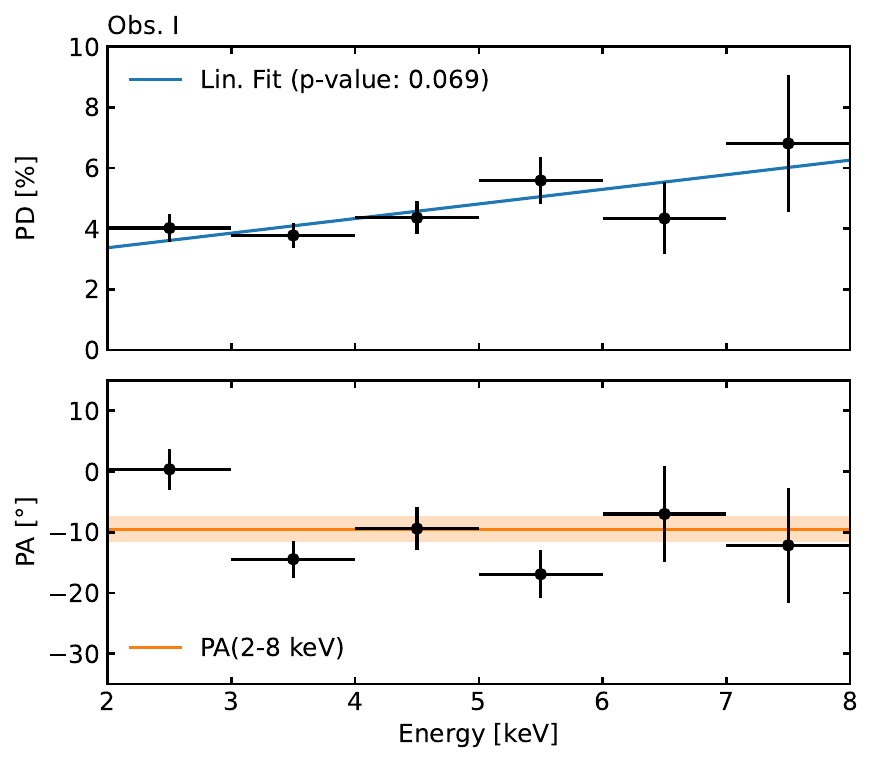}
    \end{subfigure}
    \begin{subfigure}[b]{0.2975\textwidth}
    \centering
    \caption{GX~5--1 (Obs. II)}\label{fig:PD-En.GX5-1.ObsII}
    \includegraphics[width=\textwidth]{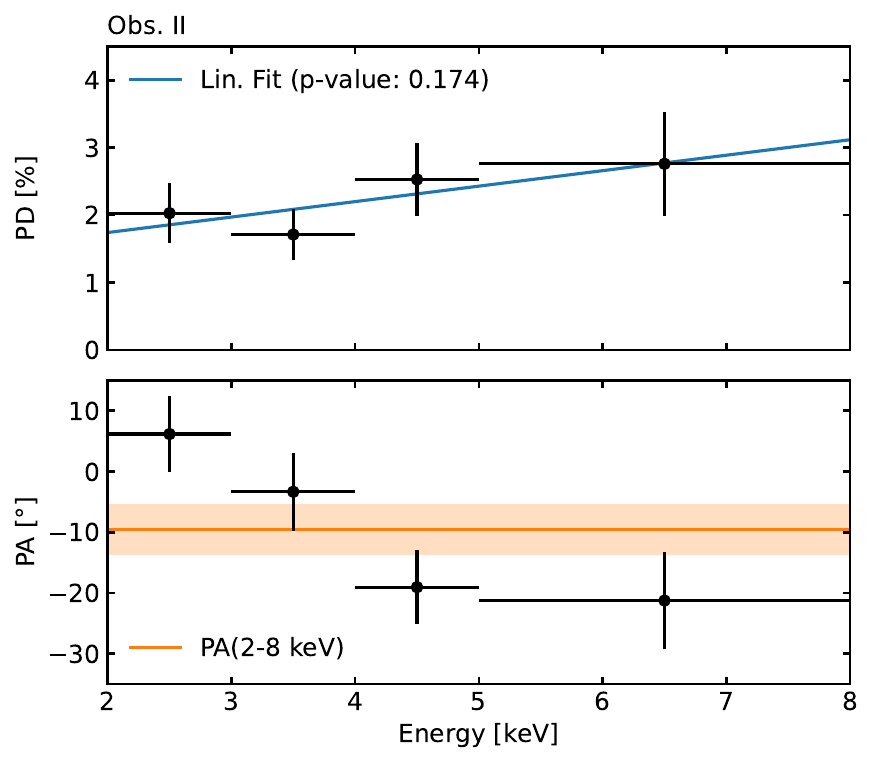}
    \end{subfigure}
    \begin{subfigure}[b]{0.2975\textwidth}
    \centering
    \caption{Sco~X-1}\label{fig:PD-En.ScoX-1}
    \includegraphics[width=\textwidth]{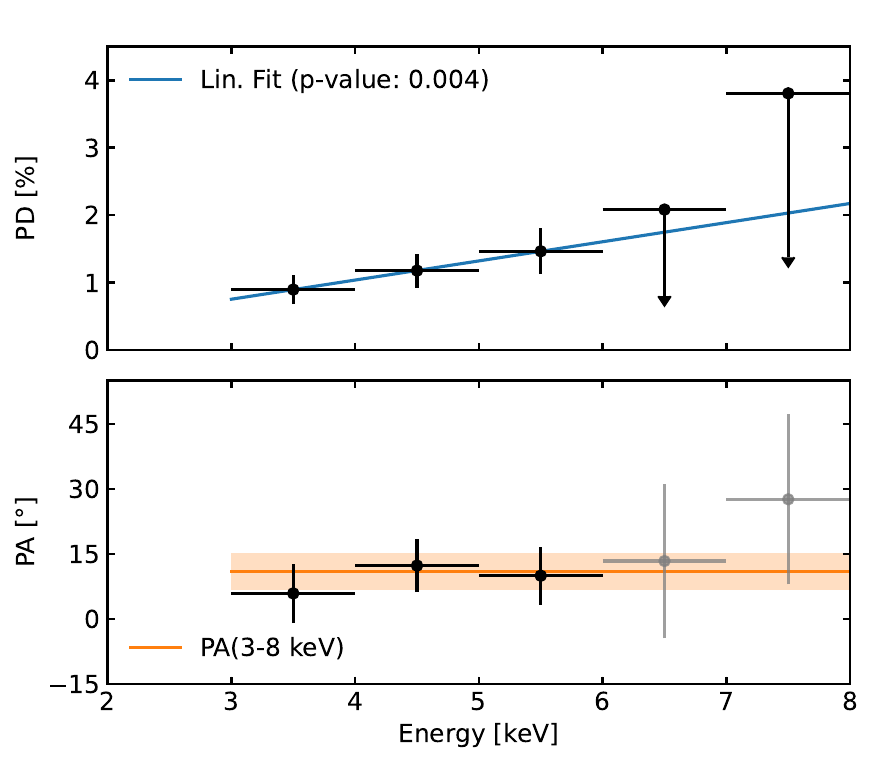}
    \end{subfigure}
    \\[1ex]
    \begin{subfigure}[b]{0.2975\textwidth}
    \centering
    \caption{GX~340+0 (Obs. I)}\label{fig:PD-En.GX340+0.ObsI}
    \includegraphics[width=\textwidth]{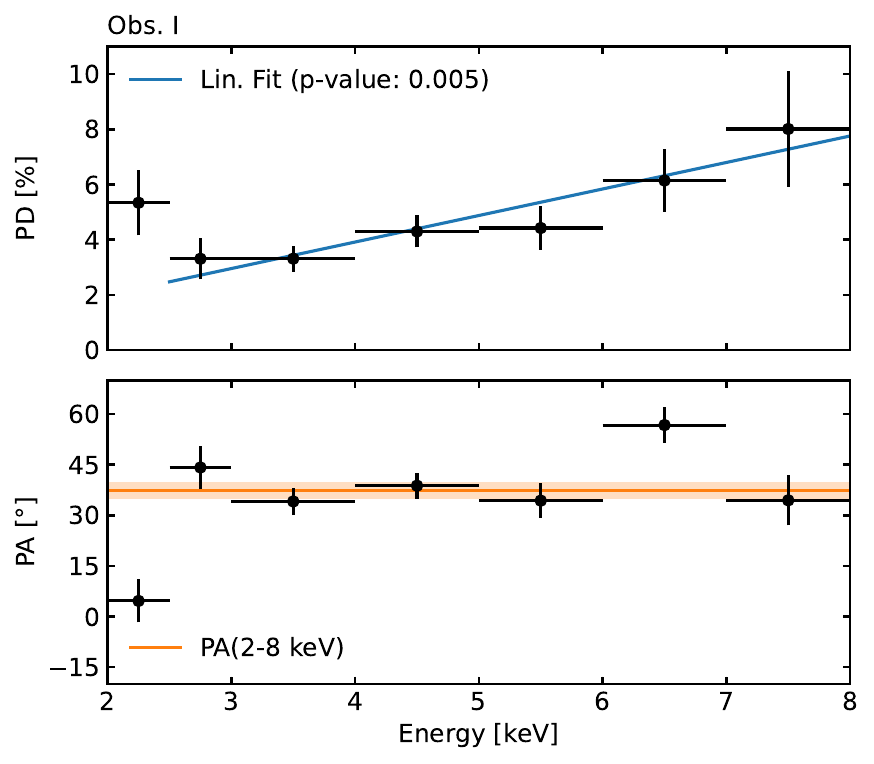}
    \end{subfigure}
    \begin{subfigure}[b]{0.2975\textwidth}
    \centering
    \caption{GX~340+0 (Obs. II)}\label{fig:PD-En.GX340+0.ObsII}
    \includegraphics[width=\textwidth]{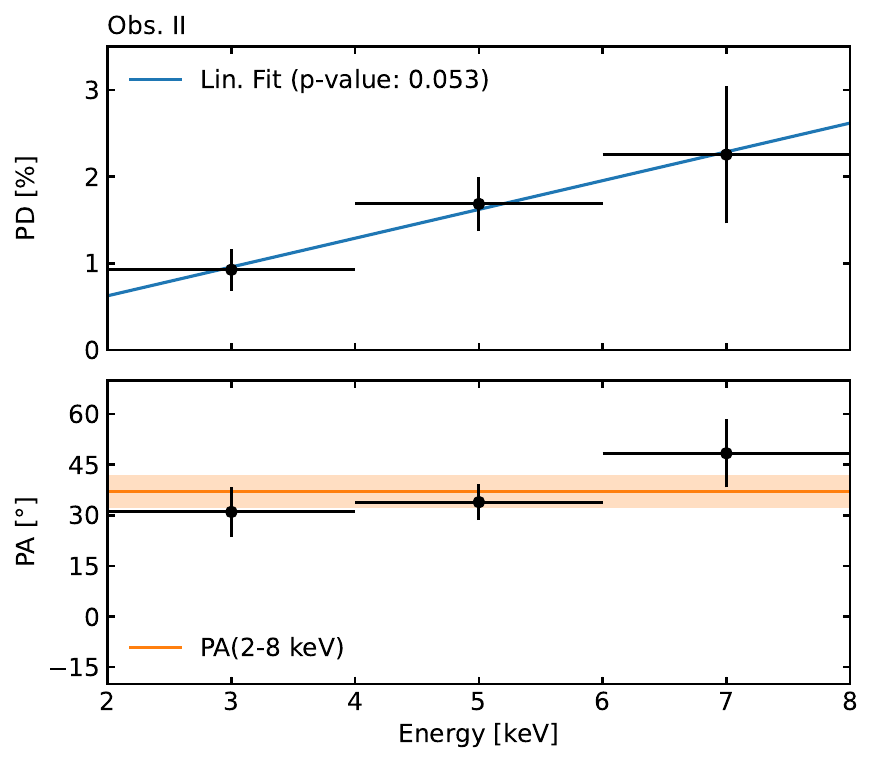}
    \end{subfigure}
    \begin{subfigure}[b]{0.2975\textwidth}
    \centering
    \caption{GX~349+2}\label{fig:PD-En.GX349+2}
    \includegraphics[width=\textwidth]{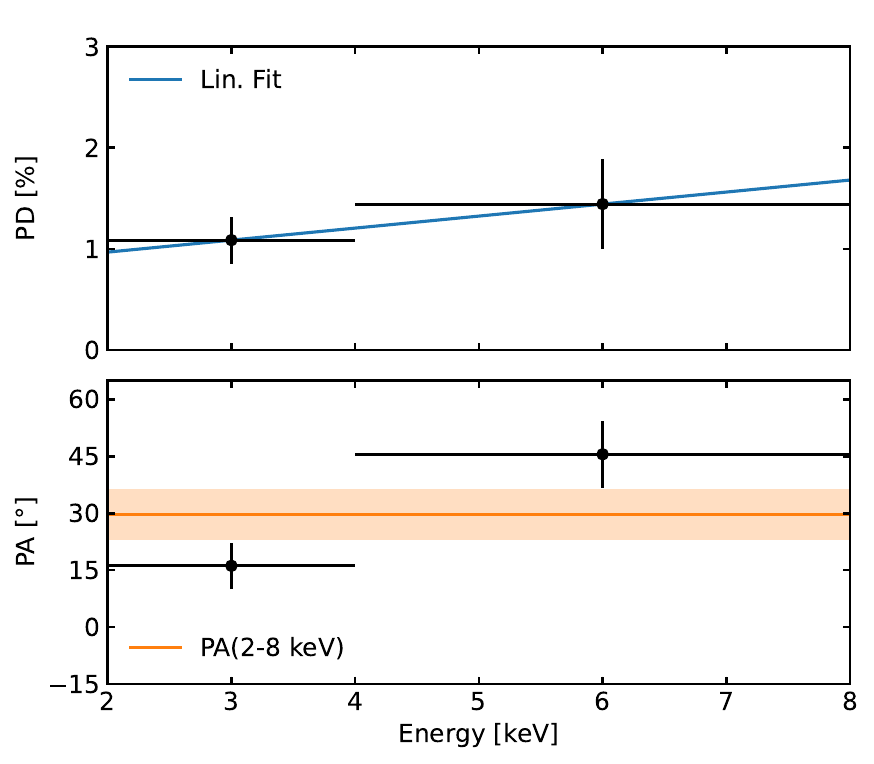}
    \end{subfigure}
    \caption{Polarization degree and angle as a function of energy. Errors are at 1$\sigma$ confidence level, while upper limits correspond to the 99\% confidence level. Blue lines show linear fits to the PD data, with resulting $p$-values. Gray points represent the PAs corresponding to the PD upper limits. Orange lines indicate the average PAs in the 2--8 keV range, with associated 1$\sigma$ error highlighted by the orange regions.}
    \label{fig:PD-En}
\end{figure*}

\subsection{Cyg~X-2}

Cyg~X-2 was the first Z-source observed by \ixpe for a total of approximately 93 ks \citep{Farinelli.etAl.2023}. However, immediately after the first observation, \ixpe observed the source again for a total of about 43 ks after the realignment of the optics modules. Until now, this observation has not been considered or analyzed. From the \ixpe CCD (Fig. \ref{fig:IXPE.CCD.CygX-2}), the source appears to remain in the same region of the Z-track (NB) during the second observation as in the first. The normalized Stokes parameters $q$ and $u$ measured by \ixpe and derived with \texttt{PCUBE} are shown in Fig. \ref{fig:Stokes.CygX-2}, along with the minimum detectable polarization (MDP99) at the 99\% confidence level. As can be easily noted, the Stokes parameters are very similar and consistent within errors between the two observations; therefore, to obtain better statistics and smaller errors for the polarization, we combined the two observations. In particular, by combining the two observations, we found a significant polarization at 6.8$\sigma$ in the 2--8 keV range (with PD = 1.8\% $\pm$ 0.2\%\footnote{All the uncertainties in the text are reported at the 1$\sigma$ confidence level} and PA = $135\degr \pm 4\degr$), compared to the significance of $5\sigma$ and $4.4\sigma$ for the two observations separately. Moreover, unlike \cite{Farinelli.etAl.2023}, by combining the two observations, we were able to study the trend of the polarization with energy in greater detail. Using 1 keV bins, the polarization can be constrained in most bins, with upper limits only in the 2--3 keV and 7--8 keV energy bands. The polarization contours were calculated for each bin using Eq. (32) of \cite{Muleri.2022} and are shown in Fig. \ref{fig:Contour.CygX-2}. The resulting PD appears to increase from lower to higher energies, up to 3.3\% $\pm$ 0.9\% in the 6--7 keV range, without any significant rotation of the PA (see also Fig. \ref{fig:PD-En.CygX-2}). We also performed a linear fit for the PD to test the significance of the increasing trend with energy (Fig. \ref{fig:PD-En.CygX-2}). The fit obtained is characterized by a $p$-value of 0.034, indicating that the observed trend is statistically significant at the 97\% confidence level. As highlighted in Sect. \ref{sec:Timing}, Cyg~X-2 tracks also the FB during the observation (see also Fig. \ref{fig:CCD.CygX-2}); therefore, we separated the data during the NB and FB. The polarization in the NB is very similar to that obtained considering the entire observations (with PD = 1.6\% $\pm$ 0.3\% and PA = $136\degr \pm 4\degr$), since the source spent most of the time in the NB. However, during the short FB interval, polarization can be constrained at 3.5$\sigma$, with a PD of 4.2\% $\pm$ 1.1\% and a PA of $125\degr \pm 7\degr$ (see also Table \ref{table:Obs}).

\subsection{XTE~J1701--462}

\ixpe observed XTE~J1701--462 twice, about ten days apart, for approximately 48 ks per observation \citep{Cocchi.etAl.2023}. During the first observation, with the source in the HB, the measured polarization is 4.4\% $\pm$ 0.4\% with $-37\degr \pm 2\degr$, well constrained at 12$\sigma$. In the second observation, the polarization drops and only an upper limit of 1.7\%\footnote{All upper limits in the text are reported at the 99\% confidence level.} is obtained. These results are consistent with those found by \cite{Cocchi.etAl.2023}. The Stokes parameters obtained with \texttt{PCUBE} are shown in Fig. \ref{fig:Stokes.XTEJ1701}. Unlike Cyg~X-2, when considering 1 keV bins, the PD significantly increases only between the first two energy bins, with a slight rotation ($\approx 20\degr$), while both the PD and PA remain almost constant for energy above 3 keV (Figs. \ref{fig:Contour.XTEJ1701.ObsI} and \ref{fig:PD-En.XTEJ1701.ObsI}). In fact, the linear fit obtained for the PD is not statistically significant in this case. However, we divided the first observation into two segments that correspond, respectively, to the upper left and upper right parts of the \ixpe CCD (Fig. \ref{fig:IXPE.CCD.XTEJ1701}), following the approach of \cite{Yu.etAl.2024}. Although the PD appears to slightly vary from 4.0\% $\pm$ 0.5\% (at 7.2$\sigma$ significance) in the softer region to 4.9\% $\pm$ 0.5\% (at 9.3$\sigma$ significance) as the source moves to the other region of the CCD, remaining consistent within the errors, the PA rotates by approximately $10\degr$. For the second observation, we obtained a significant detection only in the 5--6 keV bin, while only upper limits are found at lower and higher energies (Figs. \ref{fig:Contour.XTEJ1701.ObsII} and \ref{fig:PD-En.XTEJ1701.ObsII}). We further analyzed the second observation by  considering the three segments already introduced in Sect. \ref{sec:Timing} (see also \citealt{Zhao.etAl.2024}), and computing the average polarization in the 2--8 keV band during each segment. Although only an upper limit of 2.8\% is obtained during the second interval with the source in the FB, the polarization can be computed during the other two intervals when the source moves along the NB. During the first interval before the FB, there is a marginal detection (at 2.6$\sigma$ significance) with a PD of 2.2\% $\pm$ 0.7\% and a PA of $59\degr \pm 9\degr$, while in the interval after the FB, the PD is well detected at 3.3$\sigma$ significance with a similar PD of 2.1\% $\pm$ 0.6\% but a PA of $-57\degr \pm 8\degr$, rotated by approximately $60\degr$. In fact, summing these two intervals to compute the total polarization along the NB, we obtain only an upper limit of 2.1\%. The misalignment between the two intervals may explain this upper limit, since we are summing photons with non-parallel PAs.

\subsection{GX~5--1}

Similarly to XTE~J1701--462, \ixpe observed GX~5--1 twice, for approximately 48 ks each \citep{Fabiani.etAl.2024}. Although the time interval between the two observations is longer than that of XTE~J1701--462 ($\approx$ 20 days), \ixpe was able to cover the entire Z-track of GX~5--1 (Figs. \ref{fig:CCD.GX5-1} and \ref{fig:IXPE.CCD.GX5-1}). However, \ixpe detected significant polarization during both observations: in the first, when the source was in the HB, the polarization is 4.3\% $\pm$ 0.3\% at 13.5$\sigma$ significance, while it decreases to 2.0\% $\pm$ 0.3\% at 6.5$\sigma$ significance during the second observation, with the source moving along the NB and FB, consistent with the values obtained by \cite{Fabiani.etAl.2024}. Although the PD varies along the Z-track, the averaged PA in the 2--8 keV band remains consistent within the two observations. The normalized Stokes parameters obtained with \texttt{PCUBE} are shown in Fig. \ref{fig:Stokes.GX5-1}. If we consider the NB and FB separately during the second observation, we find a PD of 1.9\% $\pm$ 0.3\% at 5.6$\sigma$ significance in the NB, while the PD is 2.9\% $\pm$ 0.8\% during the short FB intervals (at 3.1$\sigma$ significance), without any notable rotation of the PA between the two branches. Unlike Cyg~X-2, for which the FB is significantly more polarized than the NB, in this case the PD measured in the two branches is still consistent within the errors at the 90\% confidence level. The polarization contours for the two observations are shown in Figs. \ref{fig:Contour.GX5-1.ObsI} and \ref{fig:Contour.GX5-1.ObsII}. Similarly to Cyg~X-2, the polarization is well constrained when considering 1 keV-energy bins when the source is in the HB and the polarization is higher, without any significant rotation of the PA. During the second observation, there is a slight rotation ($\approx$ 20\degr) between the first two bins at lower energies (below 4 keV) compared to the higher-energy bins (see also Figs. \ref{fig:PD-En.GX5-1.ObsI} and \ref{fig:PD-En.GX5-1.ObsII}). For the first observation, the PD increases up to 6.9\% $\pm$ 2.3\%, while the linear fit yields a $p$-value of 0.069, suggesting a possible increasing trend with energy at the 93\% confidence level. During the second observation, the PD reaches 2.8\% $\pm$ 0.8\% at high energies, but the observed increasing trend with energy is not statistically significant.

\subsection{Sco~X-1}\label{sec:ScoX-1}

Sco~X-1 was observed by \ixpe for approximately 24 ks, using the gray filter to reduce its high X-ray flux \citep{Ferrazzoli.etAl.2020,LaMonaca.etAl.2024}. The \texttt{PCUBE} algorithm of \textsc{ixpeobssim} overestimates the polarization below 3 keV when the gray filter is used, since the algorithm does not properly account for the response matrices at lower energies \citep{Veledina.etAl.2023,LaMonaca.etAl.2024}. Therefore, for this source, we computed the polarization properties in the 3--8 keV range, rather than in the full 2--8 keV band. \ixpe detected significant polarization at 6.4$\sigma$ significance in the 3--8 keV band, with a PD of 1.1\% $\pm$ 0.2\% and a PA of $11\degr \pm 4\degr$, consistent with \cite{LaMonaca.etAl.2024}. The Stokes parameters $q$ and $u$ in the 3--8 keV energy range are shown in Fig. \ref{fig:Stokes.ScoX-1}. The polarization contours are reported in Fig. \ref{fig:Contour.ScoX-1}: the PD slightly increases with energy up to 6 keV, while only an upper limit of 2.3\% is obtained in the 6--8 keV range. This increasing trend with energy is highly significant at nearly 3$\sigma$ confidence level, while the PA appears to remain constant with energy (Fig. \ref{fig:PD-En.ScoX-1}). During \ixpe observation, the source moves along the NB and the FB (see Fig. \ref{fig:CCD.ScoX-1}). In the NB, the measured PD is 0.8\% $\pm$ 0.2\% with a PA of $1\degr \pm 8\degr$ (at 3.4$\sigma$ significance), while the PD is 1.3\% $\pm$ 0.2\% with a PA of $17\degr \pm 5\degr$ (at 5.7$\sigma$ significance) along the FB. When comparing the polarization in the two branches, while the observed PD is consistent within the errors at the 90\% confidence level, the PA appears to show a slight rotation of about $15\degr$ at the same confidence level.

Throughout the entire observation and across the different branches of the Z-track, the polarization is much lower than the results obtained by PolarLight during the high-flux state, with a PA misaligned with respect to the jet direction by $\approx 50\degr$ or $\approx 35\degr$ for the NB or the FB, respectively. However, the PolarLight observation was performed by integrating over an extensive exposure time (about 884 ks; \citealt{Long.etAl.2022}), during which it was not possible to distinguish and separate the polarization of the individual Z-branches.

\subsection{GX~340+0}

\ixpe observed GX~340+0 twice \citep{LaMonaca.etAl.2024.GX340+0,Bhargava.etAl.2024a,Bhargava.etAl.2024b}. The first observation, lasting approximately 100 ks, was performed with the source moving along the HB, with significant polarization detected at 11.1$\sigma$ in the 2--8 keV (PD = 4.2\% $\pm$ 0.4\% and PA = $37\degr \pm 3\degr$). During the second, longer observation (about 190 ks), the average polarization over the entire exposure is 1.4\% $\pm$ 0.2\% with PA of $37\degr \pm 5\degr$, well constrained at 5.5$\sigma$. The normalized Stokes parameters computed with \texttt{PCUBE} are shown in Fig. \ref{fig:CCD.GX340+0}. During the first observation, the PA shows a significant rotation by about $30\degr$ between the first energy bin (i.e., 2--2.5 keV) and those at energies greater than 2.5 keV. The PD decreases between the two energy bins, from 5.3\% $\pm$ 1.2\% to 3.3\% $\pm$ 0.4\%, then increases with energy up to 8.0\% $\pm$ 2.1\% in the 7--8 keV energy bin (see Figs. \ref{fig:Contour.GX340+0.ObsI} and \ref{fig:PD-En.GX340+0.ObsI}). Excluding the first energy bin, the increasing trend with energy observed above 2.5 keV is statistically significant at the 99.5\% confidence level, while the PA appears to remain constant, with the exception of the 6--7 keV energy bin, where the Fe K$\alpha$ line is also observed. In this bin, the PA exhibits a slight rotation compared to the others but is still consistent within the errors. During the second observation, the PD increases again with energy, up to 2.3\% $\pm$ 0.8\% in the 6--8 keV range. The increasing energy trend is statistically significant around the 95\% confidence level. The PA does not show any significant rotation with energy (see Figs. \ref{fig:Contour.GX340+0.ObsII} and \ref{fig:PD-En.GX340+0.ObsII}). For most of the exposure during this observation, GX~340+0 remains in the NB with a PD of 1.5\% $\pm$ 0.3\% and a PA of $36\degr \pm 5\degr$ at 5.4$\sigma$ significance. As highlighted in Fig. \ref{fig:LC.GX340+0} by the black interval, GX~340+0 also tracks part of the HB during this observation (Fig. \ref{fig:IXPE.CCD.GX340+0}), but only an upper limit of 2.6\% is obtained \citep[see also][]{Bhargava.etAl.2024b}. This upper limit is not consistent with the results from the first observation with the source in the HB. A possible reason for this discrepancy may be a different PD along the HB: during the first observation, the source spends most of the time in the upper left part of the HB (see Fig. \ref{fig:IXPE.CCD.GX340+0}), while during the second, there is only a small excursion in that region, with most of the time spent in the right part of the HB. Therefore, it is possible that the harder region is more polarized than the softer region \citep[see also][]{Yu.etAl.2024}, for which we obtain only the upper limit, since the statistics are not sufficient to constrain the polarization. During the \ixpe observations, GX~340+0 also tracks the FB rapidly: only an upper limit of 5.1\% can be obtained for the FB, consistent with that found by \cite{Bhargava.etAl.2024b}. However, this upper limit is not very stringent and is also in line with all results obtained in the FB for other Z-sources.

\subsection{GX~349+2}

GX~349+2 was observed by \ixpe for about 96 ks, while the source was mainly in the NB with two excursions in the FB. The last FB interval was identified using the simultaneous \nustar data (Fig. \ref{fig:CCD.GX349+2}), while the others at the beginning of the observation were selected due to the similarity of the \ixpe flux behavior with that of the last FB interval. Considering the entire observation, \ixpe detected significant polarization in the 2--8 keV band at 3.8$\sigma$ significance with a PD of 1.1\% $\pm$ 0.3\% and a PA of $30\degr \pm 7\degr$. The normalized Stokes parameters derived with \texttt{PCUBE} are shown in Fig. \ref{fig:Stokes.GX349+2}. For this source, the polarization can also be constrained in the 2--4 keV and 4--8 keV energy ranges, as shown  by the polarization contours in Fig. \ref{fig:Contour.GX349+2}. We also considered smaller energy bins (e.g. 2 keV bins), but the statistics in each bin were not sufficient to constrain the polarization. A slightly increasing PD with energy and a rotation of the PA of approximately $30\degr$ between the low and high energies are observed (see also Fig. \ref{fig:PD-En.GX349+2}), similar to what was obtained during the second observation of GX~5--1. Moreover, we can compute the observation along the different branches separately: in the NB, the observed polarization is 1.0\% $\pm$ 0.3\% with a PA of $38\degr \pm 8\degr$ (at 3.3$\sigma$ significance), while the PD reaches 2.1\% $\pm$ 0.6\% (at 3.1$\sigma$ significance) when the source moves in the FB. Although the PD measured in the NB and FB is consistent within uncertainties at the 90\% confidence level, the PA differs significantly between the two branches at the same confidence level, suggesting a rotation of the PA of approximately $30\degr$. 


\section{Discussion}\label{sec:Results}

\begin{table}[t]
\caption{Polarization degree and angle for each source with the corresponding branch of the Z-track.} 
\label{table:Pol}      
\centering                                     
\begin{tabular}{l c c c}         
\hline\hline       
\noalign{\smallskip}
  Source & Branch & PD [\%] & PA [deg] \\   
\hline     
\noalign{\smallskip}
Cyg~X-2 & NB & 1.6 $\pm$ 0.3 & --44 $\pm$ 4 \\
Cyg~X-2 & FB & 4.2 $\pm$ 1.1 & --55 $\pm$ 7 \\\hline
XTE~J1701--462 & HB & 4.4 $\pm$ 0.4 & --37 $\pm$ 2 \\
XTE~J1701--462 & NB & $<2.1$ & Unconstrained \\
XTE~J1701--462 & FB & $<2.8$ & Unconstrained \\\hline
GX~5--1 & HB & 4.3 $\pm$ 0.3 & --9 $\pm$ 2 \\
GX~5--1 & NB & 1.9 $\pm$ 0.3 & --11 $\pm$ 5 \\
GX~5--1 & FB & 2.9 $\pm$ 0.8 & --6 $\pm$ 8 \\\hline
Sco~X-1\tablefootmark{$\star$} & NB & 0.8 $\pm$ 0.2 & 1 $\pm$ 8 \\
Sco~X-1\tablefootmark{$\star$} & FB & 1.3 $\pm$ 0.2 & 17 $\pm$ 5 \\\hline
GX~340+0 & HB & 4.2 $\pm$ 0.4 & 37 $\pm$ 3 \\
GX~340+0 & NB & 1.5 $\pm$ 0.3 & 36 $\pm$ 5 \\
GX~340+0 & FB & $<5.1$ & Unconstrained \\\hline
GX~349+2 & NB & 1.0 $\pm$ 0.3 & 38 $\pm$ 8 \\
GX~349+2 & FB & 2.1 $\pm$ 0.6 & 7 $\pm$ 8 \\
\hline                                            
\end{tabular}
\tablefoottext{$\star$}{Polarization is computed in the 3--8 keV range (see Sect. \ref{sec:ScoX-1}).}
\tablefoot{
Errors are quoted at the $1\sigma$ confidence level, while upper limits are reported at the 99\% confidence level for one interesting parameter.
}
\end{table}

\begin{figure}
    \centering
    \includegraphics[width=\linewidth]{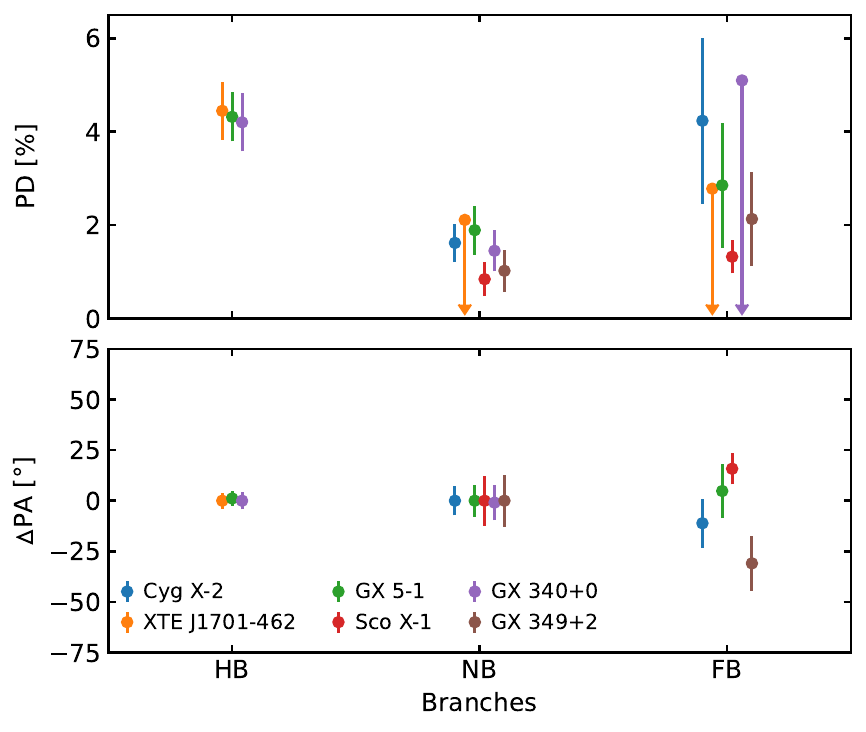}
    \caption{Polarization for each Z-source as a function of the branch (see Table \ref{table:Pol}). The polarization angle is normalized to the NB value for all sources, except for XTE~J1701--462, where the only PA measurement is in the HB. For Sco~X-1, PD and PA are computed in the 3--8 keV range (see Sect. \ref{sec:ScoX-1}). Errors are at the 90\% confidence level; upper limits are reported at the 99\% confidence level for one interesting parameter.}
    \label{fig:Pol.En}
\end{figure}

In this new analysis, we were able to constrain the polarization of Z-sources along the three different branches of their CCDs and HIDs. We summarize all the results for each source in each branch of the Z-track in Table \ref{table:Pol} and Fig. \ref{fig:Pol.En}. In Table \ref{table:Pol} we report the value of the PA as measured directly by \ixpe, which therefore depends on the orientation of the system. In Fig. \ref{fig:Pol.En} we ``renormalize'' the PA values with respect to that of the NB of each source, except for XTE~J1701--462, for which the only significant PA measurement is in the HB and only upper limits are obtained in the NB and FB. We confirm the general trend already found for some sources, characterized by decreasing polarization as they moved from the HB to the NB (see also \citealt{Cocchi.etAl.2023,Fabiani.etAl.2024}). We were able to measure the polarization for most of the Z-sources also in the FB. Whereas only upper limits were for XTE~J1701--462 and GX~340+0, for the other observed sources we found a qualitatively increasing trend of the PD moving from the NB to the FB (Table \ref{table:Pol} and Fig. \ref{fig:Pol.En}). Cyg~X-2 exhibits the greatest variation among the Z-sources, from approximately 1.5\% in the NB up to 4.2\% in the short FB intervals, without any significant rotation in the PA. GX~5--1 is the only Z-source observed along the CCD with significant polarization detected in each branch: the highest polarization is reached along the HB ($\approx 4\%$), the PD decreases  to $\approx 2\%$ in the NB and increases again in the FB up to $\approx 3\%$, while the PA remains similar between the different branches. Since in these sources the main contribution to the polarization comes from the hard Comptonization emission, the behavior of the polarization in the different branches may suggest a possible variation of the geometric configuration of the Comptonizing region along the Z-track. Slab-like configurations illuminated by the NS and the accretion disk are typically characterized by a PD that may be compatible with the results in the HB, but are too high for the NB \citep{Gnarini.etAl.2022,Gnarini.etAl.2024}. In contrast, more spherically symmetric configurations are generally low-polarized \citep{Gnarini.etAl.2022,Farinelli.etAl.2024, Bobrikova.etAl.2024.SL} and do not reach the high PD observed in the HB or in the FB. 

Therefore, there should be a change in the shape or dimension of the corona, e.g., considering a spreading or boundary layer-like geometry with different radial extension and latitudinal coverage of the NS surface as the source moves along its Z-track, varying its accretion rate. Unless there are significant spectral variations, either global or relative to individual components, detectable only through a model-dependent analysis, these polarization variations could suggest the presence of a switching mechanism for an efficient polarizer that is barely or not detectable through low-resolution spectroscopy but plays a crucial role in polarimetry. In this context, the presence of an outflowing wind could be a natural candidate. For example, \citet{Poutanen.etAl.2023}, to explain the high PD of Cyg~X-1,which is difficult to reconcile with the source inclination angle, proposed a mechanism in which the inner hot flow is outflowing with a mildly relativistic velocity $v/c \gtrsim 0.4$. Although such a high velocity almost certainly cannot be achieved in NS-LXMB winds powered by Compton heating or magnetic driving \citep{Allen.etAl.2018}, it remains true that regardless of the outflowing velocity, an optically thin wind in which photons undergo just a few scatterings could boost the polarization of the underlying system. However, a more precise evaluation of this effect requires a rather complex setup and has not yet been performed. Furthermore, unlike other Z-sources observed in the FB, Sco~X-1 and GX~349+2 also show a rotation of the PA between the NB and the FB. In particular, while the rotation of about $15\degr$ is less significant (at the 90\% confidence level) in the case of Sco~X-1, GX~349+2 exhibits a significant rotation at the 99\% confidence level of approximately $30\degr$ moving from the NB to the FB.

We also confirmed the general increasing trend of the PD with energy, already found in several NS-LXMBs, including Atoll sources (see also the review by \citealt{Ursini.2024.Review}). Moreover, some Z-sources (i.e., GX~340+0, GX~5--1, and GX~349+2) seem to exhibit a slight rotation of the PA between the lower and higher energy bins of approximately $30\degr$ (see also Figs. \ref{fig:Contours} and \ref{fig:PD-En}). This rotation of the PA with energy is not consistent with that expected considering special and general relativistic effects \citep{Connors.Stark.1977,Connors.etAl.1980,Ishihara.etAl.1988,Schnittman.2009,Schnittman.2010}. For a standard NS, depending on the geometric configuration and the inclination of the system, the rotation of the polarization plane due to relativistic effects should range between $5\degr-20\degr$. Moreover, for axisymmetric systems, we expect that the two main spectral components, i.e., the soft thermal radiation and the hard Comptonized component, should be polarized either parallel or orthogonal to each other, depending on the geometry of the Comptonizing region; therefore, the expected rotation of the PA should be $90\degr$. The observed rotation is lower, suggesting a possible misalignment between the NS spin axis and the accretion plane or a ``tilted'' inner region of the disk, breaking the symmetry of the system \citep{Abarr.2020}. It is also worth noting that an energy-dependent PA rotation of less than $90\degr$ can arise as a convolutional effect of emission at different latitudes of a spreading layer. This has recently been shown by \citet{Bobrikova.etAl.2024.SL}, who considered the case of rest-frame Chandrasekhar-like polarization with a latitudinal emission profile of the spreading layer brightness that is either uniform or follows an analytical law qualitatively matching the results of \cite{Inogamov.Sunyaev.1999}. Interestingly, in the latter case, for a reliable set of parameters, a rotation of the angle up to $40\degr$ in the \ixpe band can be achieved. However, it is important to note that this effect applies only to the spreading layer and must be combined with the contributions from both the accretion disk and reflection, meaning that the net observational result depends on the cross-normalization of each of these components.

\section{Conclusions}\label{sec:Conclusions}

The polarization properties of Z-sources are more complex than previously considered: while the HB is still the most polarized branch, exhibiting the highest PD in the entire 2--8 keV range among NS-LMXBs, the polarization first decreases along the NB and then appears to increase again in the FB. In particular, this increasing trend between the NB and FB is significant for Cyg~X-2 and Sco~X-1, while the PD value measured for GX~5--1 and GX~349+2 remain consistent at the 90\% confidence level between the two branches. Moreover, the behavior of the PA as a function of the position along the Z-track is not unique: although some sources do not show a significant rotation of the PA moving along the different branches, Sco~X-1 and GX~349+2 exhibit variations of the PA between the NB and FB at the 90\% confidence level. This rotation of the PA, as well as that observed as a function of energy for some sources, is significantly lower than $90\degr$, as would be expected for axisymmetric systems. Furthermore, the only two Z-sources (i.e., Cyg~X-2 and Sco~X-1) for which the orientation is known from the detection of the radio jet show two different behaviors: Cyg~X-2 is characterized by a PA always aligned with the radio jet direction, similar to what is observed in other classes of X-ray accreting sources such as black hole binaries or AGNs. In contrast, the PA detected for Sco~X-1 is always misaligned with respect to both the radio jet and the disk direction, likely suggesting a geometric configuration different from Cyg~X-2 or a preceding radio jet.

The spectropolarimetric analysis in the upcoming second part of this work will shed new light on the results obtained so far for the Z-sources, as it will allow us to study the polarization of each spectral component describing the X-ray emission of these sources.


\begin{acknowledgements}
This research was supported by the Italian Space Agency (Agenzia Spaziale Italiana, ASI) through the contract ASI-INAF-2022-19-HH.0 and by the Istituto Nazionale di Astrofisica (INAF) grant 1.05.23.05.06: ``Spin and Geometry in accreting X-ray binaries: The first multi frequency spectro-polarimetric campaign''. This research used data products provided by the \ixpe Team (MSFC, SSDC, INAF, and INFN) and distributed with additional software tools by the High-Energy Astrophysics Science Archive Research Center (HEASARC), at NASA Goddard Space Flight Center (GSFC). This work was supported in part by NASA through the \nicer mission and the Astrophysics Explorers Program, together with the \nustar mission, a project led by the California Institute of Technology, managed by the Jet Propulsion Laboratory, and funded by the National Aeronautics and Space Administration. The \nustar Data Analysis Software (\textsc{NUSTARDAS}), jointly developed by the ASI Science Data Center (ASDC, Italy) and the California Institute of Technology (USA), has also been used in this project. This research has made use of data and/or software provided by the High Energy Astrophysics Science Archive Research Center (HEASARC), which is a service of the Astrophysics Science Division at NASA/GSFC.
\end{acknowledgements}

\bibliographystyle{aa}
\bibliography{References}

\end{document}